\documentclass[article,nojss]{jss}

\usepackage{thumbpdf,lmodern}

\usepackage{float}
\usepackage{algorithm}
\usepackage{algpseudocode}
\usepackage{algorithmicx}
\usepackage{mathtools}
\usepackage{amsmath}
\usepackage{physics}
\usepackage{siunitx}
\usepackage{booktabs}

\MakeRobust{\Call}

\DeclareMathOperator{\erfcx}{erfcx} % Gauss scaled complementary error function
\DeclareMathOperator{\BetaFun}{B}

\defcitealias{marsaglia-tail-1}{Marsaglia's (1964)}
\defcitealias{marsaglia-ziggurat-2}{Marsaglia's and Tsang's (2000b)}

\makeatletter
\providecommand*{\input@path}{}
\g@addto@macro\input@path{{figures/}}
\makeatother
\graphicspath{{figures/}}

\author{Morteza Jalalvand\\\scriptsize{Institute for Advanced Studies in Basic Sciences}
   \And Mohammad A. Charsooghi\\\scriptsize{Institute for Advanced Studies in Basic Sciences}}
\Plainauthor{Morteza Jalalvand, Mohammad A. Charsooghi}

\title{Generalized Ziggurat Algorithm for Unimodal and Unbounded Probability Density Functions with \pkg{Zest}}
\Plaintitle{Generalized Ziggurat Algorithm for Unimodal and Unbounded Probability Density Functions with Zest}
\Shorttitle{Generalized Ziggurat Algorithm with \pkg{Zest}}

\Abstract{
We present a modified Ziggurat algorithm that could generate a random number from all unimodal and unbounded PDFs.
For PDFs that have unbounded density (e.g. gamma and Weibull with shape parameter less than one)
and/or unbounded support we use a combination of nonlinear mapping
function and rejection sampling to generate a random number from the peak and/or the tail distribution. A family of mapping
functions and their corresponding acceptance probability functions are presented (along with the criteria for their use and
their efficiency) that could be used to generate random numbers from infinite tails and unbounded densities.

The \pkg{Zest} library which is a \proglang{C++} implementation of this algorithm is also presented. \pkg{Zest} can efficiently 
generate normal, 
exponential, cauchy, gamma, Weibull, log-normal, chi-squared, student's t and Fisher's f variates.
The user can also define their custom PDF as a class and supply it as a template argument to our library's class without 
modifying 
any part of the library.
Performance of \pkg{Zest} is compared against performance of random modules of (GCC's implementation of) \pkg{Standard Template 
Library (STL)} \citep{stl} and \pkg{Boost} \citep{boost}.
The presented results show that \pkg{Zest} is faster than both in most cases, sometimes by a factor of more than 10.

We also present a \proglang{C++} implementation of a uniform floating-point random number generator (RNG) which is capable of 
producing all
representable floating-point numbers in $[0,1)$ including the denormalized numbers with correct probabilities which will be used
in the Ziggurat algorithm near unbounded peaks and tails.
The common method of dividing a random integer by the range of the RNG can not produce random floating-point numbers with fully 
random fraction bits and very small random numbers.
The presented uniform floating-point RNG is very efficient and in the case of producing double precision floating-point numbers 
it's even faster than simply multiplying a 64-bit integer by $2^{-64}$.
}

\Plainkeywords{Ziggurat algorithm, unimodal unbounded probability distributions, C++}

\Address{
  Morteza Jalalvand, Mohammad A. Charsooghi\\
  Department of Physics\\
  Institute for Advanced Studies in Basic Sciences\\
  Zanjan 45137-66731, Iran\\
  E-mail: \email{jalalvand.m@gmail.com}, \email{charsooghi@iasbs.ac.ir}\\
  URL: \url{https://iasbs.ac.ir/~jalalvand.m/}\\
  \hphantom{URL: }\url{https://iasbs.ac.ir/~charsooghi/}
}

\begin{document}

\section{Introduction}\label{sec:introduction}

Generating a random number from a given PDF has many scientific and engineering applications. Therefore, many libraries provide
facilities for generation of random number from famous PDFs. Consequently, an algorithm that is both sufficiently general and
efficient to be applicable for arbitrary distributions would be highly desirable.

Algorithms that are general enough to be applicable to an arbitrary distribution, are usually slower than algorithms specific to 
a certain distribution.
If the inverse of cumulative density function (ICDF) is available, it could be used to directly map a uniform random
number $ u \in [0,1) $ to the desired distribution \citep[chap. II, sec. 2]{devroye-book}.
But in many cases there is no closed form available for evaluation of ICDF
or it is computationally expensive. For example the ICDF of normal distribution is the inverse error function and the ICDF of 
gamma distribution is the inverse of incomplete gamma function.
\citet{uniform-ratio} presented an algorithm to generate a random number
from an arbitrary continuous probability distribution using the ratio of two uniform random numbers.
First, a point is uniformly selected from a predetermined region, then the ratio of its coordinates is returned. This is a simple
and intuitive algorithm, but in order to generate a point from desired region one usually has to use rejection methods with poor
efficiency and need to evaluate complicated functions to check if the point lies in the region or not. Although the later
requirement can be relaxed by using appropriate pretests (the so-called ``squeeze improvement'')
\citep{uniform-ratio-pretest-1,uniform-ratio-pretest-2}.
\citet{log-concave} developed a very interesting algorithm that could be used for any distribution with log-concave density with 
the rejection efficiency of exactly $\frac{1}{2}$.

Often generality could be traded for speed and algorithms tailored to a specific distribution are faster.
The Box-Muller algorithm is a fast and simple way of generating normally distributed random numbers.
The Ziggurat algorithm \citep{marsaglia-ziggurat-1,marsaglia-ziggurat-2} is an even faster generator of normally distributed 
random numbers.
There are also many efficient algorithms available for generation of gamma variates \citep{marsaglia-gamma,ahrens-gamma}.
Student's t variates could be generated with one normal and one gamma variates, and fisher's f variates could be generated given 
two gamma variates.

In this paper a generalized Ziggurat algorithm is presented which is based on the updated version of the Ziggurat algorithm 
\citep{marsaglia-ziggurat-2}. 
The Ziggurat algorithm is one of the fastest algorithms available for generation of normally distributed random 
numbers \citep{gaussian-survey}.
Although it has been stated
that Ziggurat algorithm can be used for any monotonic or symmetric unimodal PDF, lack of efficient universal algorithm to
generate random numbers from infinite tail of a general distribution renders it unusable except for the case of normal and
exponential distributions and those distributions whose ICDF can be computed.
Original Ziggurat algorithm also has a deficiency when generating random numbers from distributions with unbounded densities as
there is clearly no maximum in PDF of these distributions to place the upper edge of the topmost Ziggurat block on.

As will be shown the generalized Ziggurat algorithm is in many cases even faster than algorithms specifically developed for 
well-known distributions. It can be used for unimodal and monotone distributions with unbounded density and/or support and it 
is not limited to log-concave or light-tailed distributions.

Another motivation for providing a new Ziggurat implementation is the pattern of design flaws that can be commonly observed in
previous implementations. One such flaw that has been noted by \citet{doornik} is that the least significant bits (LSBs) of
a random integer are used both as a random index and to produce a uniform real number resulting in correlation among generated
numbers.
Another design flaw stems from the fact that unlike fixed-point numbers which have constant absolute error,
floating-point numbers have (almost) constant relative error. Values representable by a floating-point number are more closely 
spaced near
zero than they are near one. For example a 32-bit single-precision float has a precision of $2^{-149}$ near 0
while for values in $[0.5, 1)$ it is only $2^{-24}$.
The result of multiplying a 32-bit random integer by $2^{-32}$ would be a fixed-point number.
\citet{gaussian-survey}
has demonstrated how, upon conversion from fixed-point to floating-point, ``the resulting values inherit the worst of
both worlds, with lower precision near zero due to the original fixed-point value, and low
precision near one, due to the floating-point representation.''
This has serious implications for accuracy of algorithms producing random numbers from PDFs with infinite or semi-infinite 
support.
Usually log or similar functions are used in these algorithms to map values in a finite domain into an infinite range. Lack of any
non-zero number below $2^{-32}$ causes premature truncation of the tail, and the loss of precision near zero results in large gaps
in numbers produced before the truncation point.
\citet{marsaglia-float-rng} developed a new class of random number
generators specifically designed to produce floating-point values, which has later been included in
MATLAB \citep{moler-float-rng}. This algorithm is capable of producing all representable float values in $[2^{-53},1)$.
We use another algorithm based on the suggestion of \citet{gaussian-survey} to use a geometric random number
for the exponent.
This algorithm is capable of producing all representable floating-point values in $[0,1)$ including denormalized numbers with
correct probability.
It only needs about $1+2^{-9}$ 32-bit random integers per single-precision float and
$1+2^{-12}$ 64-bit random integers per double-precision float on average.
Remarkably this algorithm is even faster than the naive way in the case of double-precision floating-point numbers because it 
avoids multiplication of integers with 64 significant bits by floating-point numbers which needs a quad-precision intermediate.
Application of this algorithm for the tail distributions is well justified as it won't have any noticeable effect on the overall
speed of the Ziggurat algorithm (even in the single-precision case) but will greatly improve the accuracy of the tail 
distributions.

We hope
that this paper enhances the readers' understanding of the mechanisms used for the generation of non-uniform random numbers,
since it can provide new insights how the tail algorithms proposed by \citet{marsaglia-tail-1} and \citet{marsaglia-ziggurat-2} 
for the normal distribution
 work and achieve a high efficiency, and how and when this could be done for other
distributions, in addition to establishing clear lower bounds on the efficiency of rejection-sampling in those cases.
R. W. Hamming once said: ``The purpose of computing is insight, not numbers.''

Finally it should be mentioned that although the Ziggurat algorithm is very fast, it has a long setup times. This is not a problem
for Cauchy, normal and exponential distributions as every distribution of these kinds can be generated with shifting and scaling
the corresponding distribution with standard parameters. But for applications requiring log-normal, gamma, Weibull, student's t
or Fisher's f variates with frequently changing shape parameter, the Ziggurat algorithm is not a suitable choice.

The structure of this paper is as follows.
The original and the generalized Ziggurat algorithms are described in Section~\ref{sec:original-algorithm} and 
Section~\ref{sec:generalized-algorithm}, respectively.
The canonical uniform floating-point RNG is described in Section~\ref{sec:canonic-float-rng}.
Some of the implementation details, optimizations, and how to avoid common pitfalls and design flaws
are discussed in  in Section~\ref{sec:implementation}. Basic instructions on how to use this library can be found in 
Section~\ref{sec:usage}.
Results of tests confirming the accuracy and performance of our library are discussed in Section~\ref{sec:results}.

\section{Original Ziggurat algorithm}\label{sec:original-algorithm}

\begin{figure}[H]
 \centering{\input{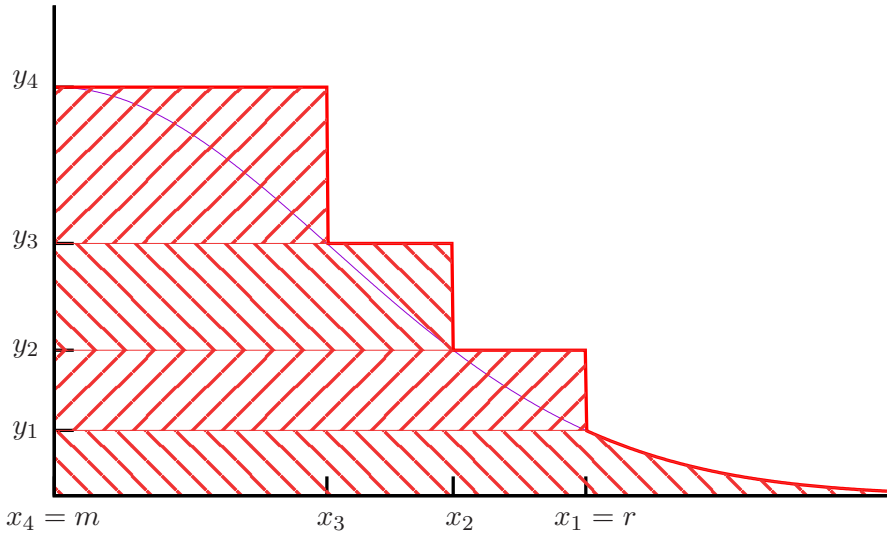}}
 \caption{The partitioning scheme of original Ziggurat algorithm with 4 regions (shown for the normal distribution). Hatched
 regions have the same area.}
 \label{fig:original-ziggurat-regions}
\end{figure}

The Ziggurat algorithm works by generating random numbers from a covering distribution that is slightly larger than desired
distribution and then rejecting those numbers that fall out of the desired distribution.
Initially, the distribution is covered with a set of $N$ equal-area regions ($N-1$ rectangles and a base strip) as shown in
Figure~\ref{fig:original-ziggurat-regions}.
The set of these regions makes up the covering distribution, whose outline is shown with the thick red line.
These regions are constructed so that
the bottom right corner of each rectangle lies on the PDF curve and the top edge of the topmost rectangle ends up at the PDF mode.
Let $f_X(x)$ denote the PDF of $X$, and $x_i$ and $y_i=f_X(x_i)$ be the coordinates of bottom right corner of the $i$th rectangle
(for $i$ between 1 and $N-1$). $x_N$ and $y_N=f_X(x_N)$ denote the top left corner of the topmost rectangle. $x_0$ should denote
the length of the rectangle with the same height and area as the base strip (i.e. $x_0 = A / y_1$ where $A$ is the area of any of
the $N$ regions).

Then to generate a random number, a region is selected with a uniform random integer $j$ in $[0,N)$. For rectangular
regions ($j \neq 0$), a uniform random real $x$ is generated in $[0,x_j)$. If $x$ is less than or equal to $x_{j+1}$,
it is inside the desired distribution and is accepted, else it will be accepted with the probability
$\flatfrac{\qty(f_X(x)-y_j)}{\qty(y_{j+1}-y_j)}$.
If $x$ is rejected, a new random region is chosen and the process will be repeated. The base strip consists of a rectangular 
region
whose area is equal to $x_1 y_1$, and a tail region; so with probability $\flatfrac{x_1 y_1}{A}$ a uniform random number within 
the
interval $[0,x_1]$ should be returned, otherwise a number from the tail region should be generated. This could be done by 
generating
a uniform random real $x$ in the interval $[0,x_0)$. $x$ will be less than or equal to $x_1$ with the probability
$\flatfrac{x_1}{x_0} = \flatfrac{x_1 y_1}{A}$, in which case it is simply returned; otherwise a number from the tail region 
should be
returned.

Algorithm~\ref{alg:original} provides the pseudocode for the original Ziggurat algorithm (some common optimizations are 
deliberately left
out for the sake of clarity). These optimizations will be discussed in Section~\ref{sec:implementation}.

\begin{algorithm}[H]
 \caption{Original Ziggurat algorithm (some common optimizations are omitted for clarity)}
 \label{alg:original}
 \begin{algorithmic}[1]
 \Require $N$ \Comment {Number of regions}
 \Require $x[0..N]$, $y[0..N]$ \Comment {Coordinates of regions}
 \Require \Call {PDF}{$x$} \Comment {Probability Density Function}
 \Require \Call {RandReal}{\null} \Comment {Random real $\in [0,1)$}
 \Require \Call {RandInteger}{$n$} \Comment {Random integer $\in [0,n)$}
 \Function{Ziggurat}{\null}
  \Loop
   \State $j \gets \Call{RandInteger}{N}$
   \State $x \gets x[j] \times \Call{RandReal}{\null}$
   \If {$x \leq x[j+1]$} \Return {$x$}
   \ElsIf {$j \not= 0$ and $\Call{RandReal}{\null}\times(y[j+1]-y[j]) < \Call{PDF}{x}-y[j]$} \Return {$x$}
   \ElsIf {$j = 0$} \Return {\Call{Tail}{$x[1]$}}
   \EndIf
  \EndLoop
 \EndFunction
 \end{algorithmic}

\end{algorithm}

To generate a number from the tail of normal distribution either Algorithm~\ref{alg:normal-tail-1} or \ref{alg:normal-tail-2}
could be used. How these algorithms work and how to do the same for other distributions will be explained in 
Section~\ref{sec:tail-algorithm}. Algorithm~\ref{alg:normal-tail-1} is a special case of Algorithm~\ref{alg:tail-ipdf} and 
Algorithm~\ref{alg:normal-tail-2} is a special case of Algorithm~\ref{alg:tail-logarithmic}. Both 
Algorithms~\ref{alg:normal-tail-1} and \ref{alg:normal-tail-2} have the same rejection efficiency. 
Algorithm~\ref{alg:normal-tail-1} requires evaluation of a square root and a logarithm, while Algorithm~\ref{alg:normal-tail-2} 
requires evaluation of two logarithms.

\begin{algorithm}[H]
 \caption{\citetalias{marsaglia-tail-1} tail function for the normal distribution}
 \label{alg:normal-tail-1}
 \begin{algorithmic}[1]
 \Require \Call {RandReal}{\null} \Comment {Random real $\in (0,1)$}
 \Function{Tail}{$s$}\Comment{$s$ is the beginning of the tail distribution}
  \Repeat
   \State $x \gets \sqrt{s^2 - 2 \ln(\Call{RandReal}{\null})}$
  \Until {$\Call{RandReal}{\null} < s/x$}
  \State \Return $x$
 \EndFunction
 \end{algorithmic}

\end{algorithm}

\begin{algorithm}[H]
 \caption{\citetalias{marsaglia-ziggurat-2} tail function for the normal distribution}
 \label{alg:normal-tail-2}
 \begin{algorithmic}[1]
 \Require \Call {RandReal}{\null} \Comment {Random real $\in (0,1)$}
 \Function{Tail}{$s$}\Comment{$s$ is the beginning of the tail distribution}
  \Repeat
   \State $x \gets - \flatfrac{\ln(\Call{RandReal}{\null})}{s}$
  \Until {$-2 \ln(\Call{RandReal}{\null}) < x^2$}
  \State \Return $s + x$
 \EndFunction
 \end{algorithmic}

\end{algorithm}

The appropriate coordinates $x_i$ and $y_i$ that creates the $N$ equal area regions has to be found by trial and error. Given an
initial guess for $x_1$ and $y_1$, the area of the base strip $A$ could be computed. The next coordinate could be computed with
this recursive relation $y_{j+1} = y_j + \frac{A}{x_j}$ and $x_j = f_X^{-1}(y_j)$ where $f_X^{-1}$ is the inverse of the
probability density function. Then deviation of $y_N$ and the true value of the probability of the mode is determined and the 
guess
will be adjusted accordingly.

\section{Generalized Ziggurat algorithm}\label{sec:generalized-algorithm}

Instead of partitioning the \emph{covering} distribution into $N$ equal area regions, we partition the \emph{original}
distribution into $N$ equal area regions ($N$ horizontal strips with equal area). When a PDF has unbounded support its base
strip would stretch to infinity horizontally (as in the original Ziggurat). Similarly, when a distribution has unbounded density
its top strip would stretch to infinity vertically. The covering distribution is the set of infinite strips plus bounding
rectangles of finite strips.

\begin{figure}[H]
 \label{fig:pdf-new-partitions}
 \centering
 \input{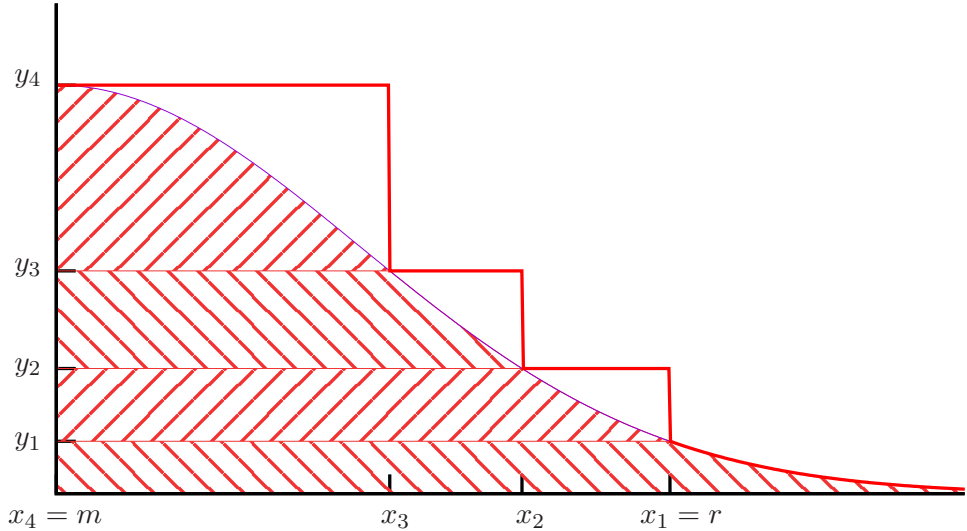}
 \caption{The partitioning scheme of generalized Ziggurat algorithm with 4 regions (shown for the normal distribution). Hatched
 regions have the same area.}
\end{figure}

With this partitioning, the first obstacle to handling a distributions with unbounded densities is overcomed. We still have to 
find
a way to generate random numbers from the infinite top strip as the method applied to the other finite strips is not applicable
here.

As it is the strips (and not their bounding rectangles) that have equal areas now, when a number is rejected instead of choosing
a new region the procedure should be repeated in the same region. Algorithm~\ref{alg:generalized} shows the pseudocode for the 
Generalized
Ziggurat algorithm.
Note that line 3 in Algorithm~\ref{alg:original} has been moved out of the loop in Algorithm~\ref{alg:generalized} so that no new 
region is selected
in case of rejection. Also note that since the condition of line 8 is always false for $j=N-1$ placing lines 3-5 after line 8
would improve the performance of the algorithm (since we don't check an unnecessary condition when $j \neq N-1$ which happens most
of the time).

\begin{algorithm}[H]
 \caption{Generalized Ziggurat algorithm}
 \label{alg:generalized}
 \begin{algorithmic}[1]
 \Require $N$ \Comment {Number of regions}
 \Require $x[0..N]$, $y[0..N]$ \Comment {Coordinates of regions}
 \Require $is\_density\_unbounded$ \Comment {Boolean}
 \Require \Call {PDF}{$x$} \Comment {Probability Density Function}
 \Require \Call {RandReal}{\null} \Comment {Random real $\in [0,1)$}
 \Require \Call {RandInteger}{$n$} \Comment {Random integer $\in [0,n)$}
 \Require \Call {Tail}{$s$} \Comment {Will be defined later}
 \Require \Call {UnboundedPeak}{$e$} \Comment {Will be defined later}
 \Function{Ziggurat}{\null}
  \State $j \gets \Call{RandInteger}{N}$
  \If {$is\_density\_unbounded$ and $j = N-1$} 
  \State\Return\Call {UnboundedPeak}{x[N-1]}
  \EndIf
  \Loop
   \State $x \gets x[j] \times \Call{RandReal}{\null}$
   \If {$x \leq x[j+1]$} \Return {$x$}
   \ElsIf {$j \not= 0$ and $\Call{RandReal}{\null}\times(y[j+1]-y[j]) < \Call{PDF}{x}-y[j]$} \Return {$x$}
   \ElsIf {$j = 0$} \Return {\Call{Tail}{$x[1]$}}
   \EndIf
  \EndLoop
 \EndFunction
 \end{algorithmic}

\end{algorithm}

\subsection{Tail algorithm}\label{sec:tail-algorithm}

In this section we discuss several algorithms that could be used to generate a random number from tail of a given distribution
with unbounded support.
Let $X$ be a random
variables with density $f_X(x)$, and let $Y$ be some function of $X$, $Y=g(X)$, with density $f_Y(y)$.
$g$ is some nonlinear mapping function that we use to map a random variable in $[0,1)$ (not necessarily uniform) into the
desired tail distribution.
Assuming $g$ is monotonic
\footnote{In case $g$ is not monotonic, the right hand side of Equation~\ref{eq:probability-change-of-var} should be a sum over 
all $x_i$
satisfying $y=g(x_i)$. For the sake of simplicity of analysis, we limit our scope to monotonic functions.}
probability contained in a differential area must be invariant under change of variable:
\begin{equation}
 \label{eq:probability-change-of-var}
 \abs\big{ f_Y(y) \dd{y} } = \abs\big{ f_X(x) \dd{x} } ,
\end{equation}
from which follows that
\begin{equation}
 \label{eq:probability-change-of-var-2}
 f_X(x) = \frac{f_Y(y)}{\abs{ \dv{x(y)}{y} }} = \frac{f_Y(y)}{\abs{ \dv{g^{-1}(y)}{y} }} .
\end{equation}

For $g$ to map $X$ in $[0,1)$ into $Y$ in tail distribution starting
at $s$, the domain of $g$ should be $[0,1)$ and its range should be $(s,+\infty)$ for the right tail or $(-\infty,s)$ for
the left tail. Along with monotonicity of $g$ this implies that there is two possible choices:
\begin{subequations}
 \begin{align}
  \label{eq:map-one-to-inf}
  \lim_{x \to 0^{+}} g(x)=s \quad\text{and}\quad \lim_{x \to 1^{-}} g(x)=\pm\infty ,
  \\
  \label{eq:map-zero-to-inf}
  \lim_{x \to 0^{+}} g(x)=\pm\infty \quad\text{and}\quad \lim_{x \to 1^{-}} g(x)=s .
 \end{align}
\end{subequations}

Equation~\ref{eq:map-one-to-inf} maps one to infinity, while Equation~\ref{eq:map-zero-to-inf} maps zero to infinity.
Theoretically, either choice does not make any difference, but from a practical point of view Equation~\ref{eq:map-zero-to-inf} is
preferable because floating-point numbers have higher precision near zero.
Of course to take full advantage of this higher precision, the random number fed into $g$ should span all the values representable
by a floating-point number. In 
Section~\ref{sec:canonic-float-rng} we present an algorithm that can produce such floating-point random numbers from
integral random numbers.

Now that we have established essential properties of the mapping function, we can discuss specific candidates and their 
properties.

\subsubsection{ICDF and ICCDF}\label{sec:tail-icdf-iccdf}

The cumulative distribution function (sometimes just called the distribution function) of a random variable $Y$ is denoted by
$F_Y(y)$ and represents the probability that the random variable $Y$ takes on values less than or equal to $y$. The CDF of a
random variable $Y$ can be expressed as the integral of its density function $f_Y$ as
\begin{equation}
 F_Y(y) = \int_{-\infty}^{y} f_Y(t) \dd{t} .
\end{equation}

The probability that a random variable $Y$ takes on values greater than $y$ is represented by the complementary cumulative
distribution function (CCDF) 
% which we denote by $\overline{F}_Y(y)$
\begin{equation}
 \overline{F}_Y(y) = \int_{y}^{+\infty} f_Y(t) \dd{t} = 1 - F_Y(y) .
\end{equation}

The CDF and CCDF are monotonic functions.
They are also strictly monotonic on the support of unimodal distributions. Therefore, their inverses
can be defined. We denote the inverse CDF (ICDF) and the inverse CCDF
(ICCDF) of $Y$ by $F_Y^{-1}(p)$ and $\overline{F}_Y^{-1}(p)$, respectively.

Substituting $F_Y^{-1}(x)$ and $\overline{F}_Y^{-1}(x)$ as mapping function $g(x)$ into 
Equation~\ref{eq:probability-change-of-var-2}
results in $f_X(x)=1$ meaning they can be used to map a uniform random variable in $[0,1)$ into $Y$.

Let $R$ and $L$ be random variables from the distribution of $Y$ conditioned by $Y>r$ and $Y<l$ respectively, so that they
correspond to the right and the left tail of the $Y$ distribution starting at $r$ and $l$, respectively.
Their densities can be found by a simple application of the Bayes theorem:
\begin{subequations}
 \begin{align}
  f_R(y) = f_{Y \,|\, Y>r}(y) =&
  \begin{dcases}
    0 & \quad\text{when}\quad y < r
    \\
%     \frac{f_Y(y)}{1-F_Y(r)} =
    \frac{f_Y(y)}{\overline{F}_Y(r)} & \quad\text{when}\quad y > r
  \end{dcases}
  \\
  f_L(y) = f_{Y \,|\, Y<l}(y) =&
  \begin{dcases}
    \frac{f_Y(y)}{F_Y(l)}
%     = \frac{f_Y(y)}{1-\overline{F}_Y(l)}
      & \quad\text{when}\quad y < l
    \\
    0 & \quad\text{when}\quad y > l .
  \end{dcases}
 \end{align}
\end{subequations}

The corresponding CDFs and CCDFs can be found by integration:
\begin{subequations}
 \label{eq:cdf-ccdf-r-l}
 \begin{align}
  F_R(y) =& \frac{F_Y(y)-F_Y(r)}{\overline{F}_Y(r)} &\quad\text{when}\quad y > r
  \\
  \overline{F}_R(y) =& \frac{\overline{F}_Y(y)}{\overline{F}_Y(r)} &\quad\text{when}\quad y > r
  \\
  F_L(y) =& \frac{F_Y(y)}{F_Y(l)} &\quad\text{when}\quad y < l
  \\
  \overline{F}_L(y) =& \frac{\overline{F}_Y(y)-\overline{F}_Y(l)}{F_Y(l)} &\quad\text{when}\quad y < l .
 \end{align}
\end{subequations}
where for the sake of brevity we have omitted the trivial cases of $y < r$ for $R$ and $y > l$ for $L$.

Inverting Equations~\ref{eq:cdf-ccdf-r-l} gives
\begin{subequations}
 \label{eq:icdf-iccdf-r-l}
 \begin{align}
  \label{eq:icdf-r}
  F_R^{-1}(p) =& F_Y^{-1} \Big( p \overline{F}_Y(r) + F_Y(r) \Big)
  \\
  \label{eq:iccdf-r}
  \overline{F}_R^{-1}(p) =& \overline{F}_Y^{-1} \Big( p \overline{F}_Y(r) \Big)
  \\
  \label{eq:icdf-l}
  F_L^{-1}(p) =& F_Y^{-1} \Big( p F_Y(l) \Big)
  \\
  \label{eq:iccdf-l}
  \overline{F}_L^{-1}(p) =& \overline{F}_Y^{-1} \Big( p F_Y(l) + \overline{F}_Y(l) \Big) .
 \end{align}
\end{subequations}

Note that Equation~\ref{eq:icdf-r} and Equation~\ref{eq:iccdf-l} satisfy Equation~\ref{eq:map-one-to-inf} while 
Equation~\ref{eq:iccdf-r} and Equation~\ref{eq:icdf-l}
satisfy Equation~\ref{eq:map-zero-to-inf},
therefore by the previously mentioned arguments the ICDF is preferable for the left tail distribution and
the ICCDF is preferable for the right tail distribution.

Algorithm~\ref{alg:tail-icdf} shows the pseudocode for generating random numbers from the left tail distribution using 
Equation~\ref{eq:icdf-l}.
The pseudocode for generating random numbers from the right tail distribution using Equation~\ref{eq:icdf-r} is similar.

\begin{algorithm}[H]
 \caption{Tail function using ICDF (suitable for left tail distributions, for right tail distributions, CDF and ICDF
 should be replaced by CCDF and ICCDF, respectively)}
 \label{alg:tail-icdf}
 \begin{algorithmic}[1]
 \Require \Call {CDF}{$x$} %\Comment {Cumulative Distribution Function}
 \Require \Call {ICDF}{$p$} %\Comment {Inverse Cumulative Distribution Function}
 \Require \Call {RandReal}{\null} \Comment {Random real $\in [0,1)$}
 \Function{Tail}{$s$}\Comment{$s$ is the beginning of the tail distribution}
  \State \Return {\Call{ICDF}{$\Call{RandReal}{\null} \times \Call{CDF}{s}$}}
 \EndFunction
 \end{algorithmic}

\end{algorithm}

\subsubsection{IPDF}\label{sec:tail-ipdf}

Unimodal probability density functions (PDF) can be broken down into two monotonic functions on either side of the mode.
If it's also strictly monotonic
\footnote{This is true for all well-known distributions}, the inverse PDF (IPDF) can be defined.
It's easy to verify that $f_Y^{-1}(x f_Y(s))$ satisfies Equation~\ref{eq:map-zero-to-inf}.
Substituting $y = f_Y^{-1}(x f_Y(s))$ as mapping function $g(x)$ into Equation~\ref{eq:probability-change-of-var-2} gives
\begin{equation}
 \label{eq:ipdf-prob-x}
 f_X(x)=\frac{f_Y(y)}{\pm\frac{1}{f_Y(s)} f'_Y(y)} = \pm x \frac{\qty[f_Y(s)]^2}{f'_Y(y)} ,
\end{equation}
where the upper sign (plus in this case) corresponds to the left tail distribution and the lower sign (minus in this case)
corresponds to the right tail distribution.

To generate $X$ with such distribution we use rejection sampling. The acceptance probability should be proportional to
the right hand side of Equation~\ref{eq:ipdf-prob-x},
\begin{equation}
 \label{eq:ipdf-accept-prob-propto}
 \Pr(x) \propto \frac{f_Y(y)}{\pm f'_Y(y)} \propto \frac{x}{\pm f'_Y(y)} .
\end{equation}

If
\begin{equation}
 \label{eq:ipdf-monotonicity-condition}
 \dv{y}(\frac{f_Y(y)}{f'_Y(y)}) 
 = 1 - \frac{f_Y(y) f''_Y(y)}{\qty[f'_Y(y)]^2}
  > 0 ,
\end{equation}
then the right hand side of Equation~\ref{eq:ipdf-prob-x} is a monotonically increasing function for the left tail or a 
monotonically
decreasing function for the right tail, therefore,
its maxima would be at the start of the tail $s$. Setting the proportionality constant such that
$\Pr(g^{-1}(s)=1)=1$ results in
\begin{equation}
 \label{eq:ipdf-accept-prob}
 \Pr(x) = \frac{f_Y(y)}{f_Y(s)}\frac{f'_Y(s)}{f'_Y(y)}
  = x \frac{f'_Y(s)}{f'_Y(y)} .
\end{equation}

Note that the condition of Equation~\ref{eq:ipdf-monotonicity-condition} can also be written in this form:
\begin{equation}
 \dv{y}(\frac{f'_Y(y)}{f_Y(y)}) = \dv[2]{y}\qty\Big(\log(f_Y(y))) < 0 ,
\end{equation}
which is the well known condition of log-concavity \citep{log-concave,devroye-book}. Many famous PDFs are log-concave, however, 
there are several notable exceptions. We show later in this section how this condition can be relaxed in those cases.

Algorithm~\ref{alg:tail-ipdf} shows the pseudocode for generating random numbers from the tail distribution using the IPDF.

\begin{algorithm}[H]
 \caption{Tail function using IPDF}
 \label{alg:tail-ipdf}
 \begin{algorithmic}[1]
 \Require \Call {PDF}{$x$} %\Comment {Cumulative Distribution Function}
 \Require \Call {IPDF}{$p$} %\Comment {Inverse Cumulative Distribution Function}
 \Require \Call {Derivative}{$x$} \Comment {Derivative of the PDF}
 \Require \Call {RandReal}{\null} \Comment {Random real $\in [0,1)$}
 \Function{Tail}{$s$}\Comment{$s$ is the beginning of the tail distribution}
  \Repeat
   \State $u_1 \gets \Call{RandReal}{\null}$
   \State $x \gets \Call {IPDF}{u_1 \times \Call {PDF}{s}}$
  \Until 
   {$\Call{RandReal}{\null} < \frac{\Call {PDF}{x}}{\Call {PDF}{s}} \times \frac{\Call {Derivative}{s}}{\Call {Derivative}{x}}$}
  \Comment {$\frac{\Call {PDF}{x}}{\Call {PDF}{s}}$ can be replaced by $u_1$}
  \State \Return {$x$}
 \EndFunction
 \end{algorithmic}

\end{algorithm}

Only ratios of PDF and PDF derivative appear in Equation~\ref{eq:ipdf-accept-prob} and Algorithm~\ref{alg:tail-ipdf}. Usually 
they can be
simplified considerably, improving both performance and accuracy. For example in the case of normal distribution
$f_Y(y)=\frac{1}{\sqrt{2\pi}\sigma}\exp(-\frac{(y-\mu)^2}{2\sigma^2})$, Equation~\ref{eq:ipdf-accept-prob} simplifies to
\begin{equation}
 \label{eq:normal-accept-prob}
 \Pr(x) = \frac{f_Y(y)}{f_Y(s)}\frac{f'_Y(s)}{f'_Y(y)} = \frac{s-\mu}{y-\mu} ,
\end{equation}
where
\begin{equation}
 \label{eq:normal-mapping-func}
 y=g(x)=f_Y^{-1}\big(x f_Y(s)\big)=\mu+\sqrt{(s-\mu)^2 - 2 \sigma^2 \ln(x)} .
\end{equation}

This is the same as \citetalias{marsaglia-tail-1} tail algorithm for the normal 
distribution provided that $\mu=0$ and
$\sigma=1$ (corresponding to the standard normal distribution).

Intuitively what's happening here is that we generate random numbers from a covering distribution whose density is proportional to
the derivative of desired PDF, hence the ICDF of the covering distribution is related to the IPDF of the desired distribution.
Log-concavity is what makes the decay of density of covering distribution slower than that of the desired distribution so that a
rejection-sampling could be performed.

Integrating Equation~\ref{eq:ipdf-accept-prob} over x in $[0,1]$ yields the efficiency of the rejection sampling:
\begin{equation}
 \label{eq:ipdf-efficiency}
 \eta =
 \begin{dcases}
  \frac{F_Y(s) \abs{f'_Y(s)}}{\qty\big[f_Y(s)]^2} & \quad\text{for left tail}
  \\
  \frac{\overline{F}_Y(s) \abs{f'_Y(s)}}{\qty\big[f_Y(s)]^2} & \quad\text{for right tail}
 \end{dcases}
\end{equation}
Observe that if the tail distribution is convex, a right triangle with height $f_Y(s)$ and base
$\flatfrac{f_Y(s)}{\abs{f'_Y(s)}}$ could be completely contained inside the tail distribution, hence there's a lower
bound of $\frac{1}{2}$ for the efficiency of this method.

Using the above formulas the efficiency of Marsaglia's tail algorithm for the standard normal distribution would be
$\sqrt{\frac{\pi}{2}} s \erfcx\qty(\frac{s}{\sqrt{2}})$, which yields 65.57\% and 91.38\% for $s=1$ and $s=3$, respectively.
These are close to 66\% and 88\% values calculated by \citet{marsaglia-tail-1}.

There are still two limitations. One is the assumption of log-concavity, which will be addressed in the following.
The other is that many distributions' IPDF is not easily computable, which will be addressed in Section~\ref{sec:tail-iipdf}.

This condition of log-concavity can be somewhat relaxed but can not be entirely eliminated.

Using $y = f^{-1}(x^\alpha f(s))$ as the mapping function $g(x)$ where $\alpha$ is a positive exponent, changes
Equation~\ref{eq:ipdf-prob-x}
into
\begin{equation}
 \label{eq:relaxed-ipdf-prob-x}
 f_X(x)=\frac{f_Y(y)}{\pm\qty[\frac{f_Y(y)}{f_Y(s)}]^{\frac{1}{\alpha}-1}\frac{1}{f_Y(s)}f'_Y(y)} \propto
 \frac{\qty[f_Y(y)]^{2-\frac{1}{\alpha}}}{\pm f'_Y(y)}
 \propto \frac{x^{2 \alpha - 1}}{\pm f'_Y(y)} ,
\end{equation}
and therefore,
\begin{equation}
 \label{eq:relaxed-ipdf-accept-prob-propto}
 \Pr(x) \propto \frac{\qty[f_Y(y)]^{2-\frac{1}{\alpha}}}{\pm f'_Y(y)}
 \propto \frac{x^{2 \alpha - 1}}{\pm f'_Y(y)} ,
\end{equation}
then the condition of Equation~\ref{eq:ipdf-monotonicity-condition} would be
\begin{equation}
 \label{eq:relaxed-ipdf-monotonicity-condition}
 \dv{y}(\frac{\qty[f_Y(y)]^{2-\frac{1}{\alpha}}}{f'_Y(y)}) =
  \qty[f_Y(y)]^{1-\frac{1}{\alpha}} \qty[2 - \frac{1}{\alpha} - \frac{f_Y(y) f''_Y(y)}{\qty[f'_Y(y)]^2}] > 0 ,
\end{equation}
and finally the acceptance probability becomes
\begin{equation}
 \label{eq:relaxed-ipdf-accept-prob}
 \Pr(x) = \qty[\frac{f_Y(y)}{f_Y(s)}]^{2-\frac{1}{\alpha}}\frac{f'_Y(s)}{f'_Y(y)}
  = x^{2 \alpha - 1} \frac{f'_Y(s)}{f'_Y(y)} .
\end{equation}

Integrating Equation~\ref{eq:relaxed-ipdf-accept-prob} gives the same efficiencies as the Equation~\ref{eq:ipdf-efficiency} 
multiplied by a factor
of $\frac{1}{\alpha}$, which suggests that the lowest possible value for $\alpha$ should be selected to optimize efficiency.
Also by the same arguments, a lower bound of $\frac{1}{2\alpha}$ holds for the efficiency.

As an example let's consider the student's t distribution:
$$f_Y(y) = \frac{1}{\sqrt{\nu} \BetaFun\qty(\frac{\nu}{2},\frac{1}{2})}
\qty(1+\frac{y^2}{\nu})^{-\frac{\nu+1}{2}} ,$$
where $\nu$ is a positive degrees of freedom parameter. This distribution is not log-concave but satisfies
Equation~\ref{eq:relaxed-ipdf-monotonicity-condition} with $\alpha \geq \frac{\nu+1}{\nu}$.
Therefore to produce a random number from the tail of student's t distribution (starting at $s$), one should set
\begin{equation}
 \label{eq:student-t-mapping-func}
 y = f_Y^{-1}(x^\alpha f_Y(s)) = \sqrt{x^{\frac{-2}{\nu}}\qty(\nu + s^2) - \nu} ,
\end{equation}
and accept it with probability
\begin{equation}
 \label{eq:student-t-accept-prob}
 \Pr(x) = \frac{s}{y} \sqrt{\frac{1 + \frac{y^2}{\nu}}{1 + \frac{s^2}{\nu}}} 
   = \sqrt{\frac{1 + \frac{\nu}{y^2}}{1 + \frac{\nu}{s^2}}} .
\end{equation}

In the $\nu \to \infty$ limit, Equations~\ref{eq:student-t-mapping-func} and \ref{eq:student-t-accept-prob} tend to
Equations \ref{eq:normal-mapping-func} and \ref{eq:normal-accept-prob}, respectively, as they should.

\subsubsection{Incomplete IPDF}\label{sec:tail-iipdf}

As mentioned in the previous section the IPDF of many distributions are not easily computable.
The inverse of PDFs that are product of an algebraic expression of $y$ and an exponential or logarithmic function of $y$, can only
be expressed in terms of the Lambert-W function
\footnote{Some examples of this kind are the chi-squared, gamma, Weibull, log-normal, log-cauchy, and l\'evy distributions.}.
Some PDFs are in the form of ratio of two algebraic expressions, each of which invertible, but whose ratio is not
\footnote{Some examples of this kind are the Burr, and log-logistic distributions.}.
In all of these cases a part of the PDF that monotonically tends to zero as $y$ approaches infinity, can be factored
(for example the exponential function or the denominator of the ratio). We shall call this the decaying part. In other words,
we assume $f_Y(y)=d(y) r(y)$ where $d(y)$ is the monotonic decaying part that tends to zero as $y$ tends to infinity,
and $r(y)$ is all the other factors of the PDF. Only the inverse of the decaying part of the PDF is used as the mapping
function $y = d^{-1}\big(x \, d(s)\big)$ which again satisfies Equation~\ref{eq:map-zero-to-inf}. Since this does not invert the 
PDF
completely, we call it the incomplete IPDF (IIPDF). Substituting this into Equation~\ref{eq:probability-change-of-var-2} yields
\begin{equation}
 \label{eq:iipdf-prob-x}
 f_X(x)=\frac{f_Y(y)}{\pm\frac{1}{d(s)}d'(y)} ,
\end{equation}
where the signs has the same role as in Equation~\ref{eq:ipdf-prob-x}. Again assuming,
\begin{equation}
 \label{eq:iipdf-monotonicity-condition}
 \dv{y}(\frac{f_Y(y)}{d'(y)}) > 0 ,
\end{equation}
the acceptance probability at the start of the tail $s$ could be set to unity resulting in
\begin{equation}
 \label{eq:iipdf-accept-prob}
 \Pr(x) = \frac{f_Y(y)}{f_Y(s)}\frac{d'(s)}{d'(y)} .
\end{equation}

The pseudocode for the tail function using the IIPDF is analogous to Algorithm~\ref{alg:tail-ipdf}. IPDF should be replaced by 
IIPDF and
the derivative function is the derivative of just the decaying part. The comment in line 5, however, does not apply any more.

The other limitation that needs to be addressed is the condition of log-concavity and its analogue for IIPDF
Equation~\ref{eq:iipdf-monotonicity-condition}.

The constraint of Equation~\ref{eq:iipdf-monotonicity-condition} can be relaxed in a manner similar to that of the previous 
section by
using $d^{-1}\big(x^\alpha  d(s)\big)$ asthe mapping function $g(x)$.

\subsubsection{Logarithmic}\label{sec:tail-logarithmic}

The method described in this section can be applied to any distribution provided that the tail of their PDF can be bound by some
exponential function (i.e., the density must be light-tailed \citep{heavy-tail-book}). First an exponential variate is generated,
then rejection-sampling is used to get the desired distribution.

The general form of the logarithmic mapping function is
\begin{equation}
 y = g(x) = s \pm \sigma \ln(x) ,
\end{equation}
where $s$ is the start of the tail distribution, $\sigma$ is a positive scale parameter, and the upper sign (plus in this case)
corresponds to left tail distributions and the lower sign (minus in this case) corresponds to right tail distributions.
Substituting this into Equation~\ref{eq:probability-change-of-var-2}
gives
\begin{equation}
 f_X(x) = \sigma f_Y(y) \exp(\mp \frac{y - s}{\sigma}) = \sigma \frac{f_Y(y)}{x} ,
\end{equation}
suggesting
\begin{equation}
 \label{eq:logarithmic-accept-prob-propto}
 \Pr(x) \propto f_Y(y) \exp(\mp \frac{y - s}{\sigma}) = \frac{f_Y(y)}{x} ,
\end{equation}
where $\sigma$ must be chosen such that
\begin{equation}
 \label{eq:logarithmic-monotonicity-condition}
 \dv{y}(f_Y(y) \exp(\frac{y - s}{\sigma})) < 0 ,
\end{equation}
which ensures that the maximum of Equation~\ref{eq:logarithmic-accept-prob-propto} occurs at $s$. This is only possible for 
light-tailed
distributions since heavy-tailed distributions can not be bound by any exponential function.
Setting $\Pr(g^{-1}(s)=1)=1$ results in
\begin{equation}
 \label{eq:logarithmic-accept-prob}
 \Pr(x) = \frac{f_Y(y)}{f_Y(s)} \exp(\mp \frac{y - s}{\sigma}) = \frac{1}{x}\frac{f_Y(y)}{f_Y(s)} .
\end{equation}

Algorithm~\ref{alg:tail-logarithmic} shows the pseudocode for tail function using a logarithmic mapping function.

\begin{algorithm}[H]
 \caption{Tail function with logarithmic mapping function (suitable for light-tailed PDFs)}
 \label{alg:tail-logarithmic}
 \begin{algorithmic}[1]
 \Require $\sigma$ \Comment {Positive real satisfying Equation~\ref{eq:logarithmic-monotonicity-condition}}
 \Require \Call {PDF}{$x$} %\Comment {Cumulative Distribution Function}
 \Require \Call {RandReal}{\null} \Comment {Random real $\in [0,1)$}
 \Function{Tail}{$s$}\Comment{$s$ is the beginning of the tail distribution}
  \Repeat
   \State $u_1 \gets \Call{RandReal}{\null}$
   \State $x \gets s \pm \sigma \ln(u_1)$
  \Until 
   {$u_1 \times \Call{RandReal}{\null} < \frac{\Call {PDF}{x}}{\Call {PDF}{s}}$}
  \State \Return {$x$}
 \EndFunction
 \end{algorithmic}

\end{algorithm}

Rejection efficiency of this method can be obtained easily by integrating Equation~\ref{eq:logarithmic-accept-prob}. It would be
$\frac{1}{\sigma}\frac{F_Y(s)}{f_Y(s)}$ for left tail distributions, and $\frac{1}{\sigma}\frac{\overline{F}_Y(s)}{f_Y(s)}$ for
right tail distributions. Therefore, $\sigma$ should be the smallest value satisfying 
Equation~\ref{eq:logarithmic-monotonicity-condition}
in order to maximize the efficiency.

It can be shown that in the case of standard normal distribution, this algorithm is equivalent to 
Algorithm~\ref{alg:normal-tail-2} proposed by
\citet{marsaglia-ziggurat-2} with $\sigma = \frac{1}{s}$ after some simplification. Note that
$\sigma = \frac{1}{s}$
is the smallest value satisfying Equation~\ref{eq:logarithmic-monotonicity-condition} for the tail of standard normal 
distribution.

\subsubsection{Trigonometric}\label{sec:tail-trigonometric}

For PDFs that are heavy-tailed but decay at least as fast as the Cauchy distribution, one can use the Cauchy distribution as the
covering distribution and generate variates from tail of the Cauchy distribution using Equations~\ref{eq:iccdf-r} and 
\ref{eq:icdf-l} and 
then use
rejection-sampling to generate the desired variate.
The mapping function would be
\begin{equation}
 g(x) = \gamma \tan(x \qty[\atan(\frac{s-y_0}{\gamma}) \pm\frac{\pi}{2}] \mp \frac{\pi}{2}) + y_0 .
\end{equation}
Here $\gamma$ and $y_0$ are the scale and location parameter of the covering Cauchy distribution and the sign convention is the
same as previous section.

With a procedure similar to that of previous section, one can show that $\gamma$ and $y_0$ should satisfy
\begin{equation}
 \label{eq:trigonometric-monotonicity-condition}
 \mp\dv{y}(f_Y(y) \qty(1 + \qty(\frac{y - y_0}{\gamma})^2)) < 0 ,
\end{equation}
and the acceptance probability would be
\begin{equation}
 \label{eq:trigonometric-accept-prob}
 \Pr(x) = \frac{f_Y(y)}{f_Y(s)} \frac{1 + \qty(\frac{y - y_0}{\gamma})^2}{1 + \qty(\frac{s - y_0}{\gamma})^2} .
%  = \frac{f_Y(y)}{f_Y(s)} \frac{\csc[2](x \qty[\atan(\frac{s-y_0}{\gamma}) \mp\frac{\pi}{2}])}{1 + \qty(\frac{s - 
% y_0}{\gamma})^2}
\end{equation}

The pseudocode for this mapping function is shown in Algorithm~\ref{alg:tail-trigonometric}.

\begin{algorithm}[H]
 \caption{Tail function with trigonometric mapping function}
 \label{alg:tail-trigonometric}
 \begin{algorithmic}[1]
 \Require $\gamma$ \Comment {Positive real satisfying Equation~\ref{eq:trigonometric-monotonicity-condition}}
 \Require $y_0$ \Comment {Real satisfying Equation~\ref{eq:trigonometric-monotonicity-condition}}
 \Require \Call {PDF}{$x$} %\Comment {Cumulative Distribution Function}
 \Require \Call {RandReal}{\null} \Comment {Random real $\in [0,1)$}
 \Function{Tail}{$s$}\Comment{$s$ is the beginning of the tail distribution}
  \State $C_1 \gets \atan(\frac{s-y_0}{\gamma}) \mp\frac{\pi}{2}$
  \State $C_2 \gets 1 + \qty(\frac{s - y_0}{\gamma})^2$
  \Repeat
%    \State $u_1 \gets \Call{RandReal}{\null}$
   \State $t \gets \tan(C_1 \times \Call{RandReal}{\null} \pm \frac{\pi}{2})$
   \State $x \gets \gamma t + y_0$
  \Until {$C_2 \times \Call{RandReal}{\null} < \qty(1 + t^2) \times \frac{\Call {PDF}{x}}{\Call {PDF}{s}}$}
  \State \Return {$x$}
 \EndFunction
 \end{algorithmic}

\end{algorithm}

\subsubsection{Rational}\label{sec:tail-rational}

The mapping function presented in this section can be used for any distribution that could be bound by a function of the form
$\abs{y}^{-(\alpha+1)}$ with $\alpha>0$. This is a (decaying) exponent function whose power is less than $-1$.
So this mapping function could be applied to almost any heavy-tailed distribution.
A type II Pareto distribution starting at $s$ is used as the covering distribution. The general form of the rational mapping 
function is
\begin{equation}
 g(x) = s \mp \sigma \qty(x^{-\qty(\alpha^{-1})} - 1) ,
\end{equation}
where $\alpha$ is a positive shape parameter and $\sigma$ is a positive scale parameter and the sign convention is the same as
previous sections.

Following a procedure similar to that of previous sections, it can be shown that $\alpha$ and $\sigma$ must satisfy
following equation for the Pareto distribution to actually cover the desired distribution
\begin{equation}
 \label{eq:rational-monotonicity-condition}
 \mp\dv{y}(f_Y(y) \qty(1 \mp \frac{y - s}{\sigma})^{\alpha+1}) < 0 ,
\end{equation}
and the acceptance probability would be
\begin{equation}
 \label{eq:rational-accept-prob}
 \Pr(x) = \frac{f_Y(y)}{f_Y(s)} \qty(1 \mp \frac{y-s}{\sigma})^{\alpha+1}
   = \frac{f_Y(y)}{f_Y(s)} x^{-\qty(1+\alpha^{-1})} .
\end{equation}

Algorithm~\ref{alg:tail-rational} shows the pseudocode for the tail distribution using the rational mapping function.

\begin{algorithm}[H]
 \caption{Tail function with rational mapping function}
 \label{alg:tail-rational}
 \begin{algorithmic}[1]
 \Require $\alpha$ \Comment {Positive real satisfying Equation~\ref{eq:rational-monotonicity-condition}}
 \Require $\sigma$ \Comment {Positive real satisfying Equation~\ref{eq:rational-monotonicity-condition}}
 \Require \Call {PDF}{$x$}
 \Require \Call {RandReal}{\null} \Comment {Random real $\in [0,1)$}
 \Function{Tail}{$s$}\Comment{$s$ is the beginning of the tail distribution}
  \Repeat
   \State $u_1 \gets \Call{RandReal}{\null}$
   \State $t \gets u_1^{-(\alpha^{-1})}$
   \State $x \gets s \mp \sigma \qty(t - 1)$
  \Until {$u_1 \times \Call{RandReal}{\null} < t \times \frac{\Call {PDF}{x}}{\Call {PDF}{s}}$}
  \State \Return {$x$}
 \EndFunction
 \end{algorithmic}

\end{algorithm}

By integrating Equation~\ref{eq:rational-accept-prob},
the efficiency of rejection-sampling could be easily found to be
$ \frac{\alpha}{\sigma} \frac{F_Y(s)}{f_Y(s)} $
and
$ \frac{\alpha}{\sigma} \frac{\overline{F}_Y(s)}{f_Y(s)}$
for right and left tail distributions, respectively.

\subsubsection{Exponential}\label{sec:tail-exponential}

Rarely distributions have a logarithmic decay  (some authors refer to these distributions as super heavy-tailed
\citep{super-heavy-1,super-heavy-2}) and can not be
bound by any exponential or power-law functions. Therefore, methods presented in
Sections~\ref{sec:tail-logarithmic}, \ref{sec:tail-trigonometric}, and \ref{sec:tail-rational}
would not be applicable. The only well known distributions with logarithmic decay are the log-Cauchy and the log-Pareto
distributions.
This mapping function would be applicable to any distribution that could be bound by a function of the form
$\abs{y}^{-1}\qty(ln(\abs{y}))^{-\alpha}$ for $\alpha>0$.
% This includes all well-known distributions.

% Using a log-Pareto distribution starting at $s$ as the covering distribution,
% the mapping function would be
The general form of the exponential mapping function is as follows
\begin{equation}
 g(x) = s \mp \sigma \qty[\exp( x^{-\qty(\alpha^{-1})}-1) - 1] ,
\end{equation}
where $\alpha$ is a positive shape parameter and $\sigma$ is a positive scale parameter. The sign convention is the same as
previous sections.

By following a procedure similar to that of previous sections, the constraint on $\alpha$ and $\sigma$ could be found as
\begin{equation}
 \label{eq:exponential-monotonicity-condition}
 \mp\dv{y}(f_Y(y)\qty(1 \mp \frac{y - s}{\sigma}) \qty(1 + \ln(1 \mp \frac{y - s}{\sigma}))^{\alpha+1}) < 0 ,
\end{equation}
and the acceptance probability would be
\begin{align}
 \label{eq:exponential-accept-prob}
 \Pr(x) & = \frac{f_Y(y)}{f_Y(s)} \qty(1 \mp \frac{y - s}{\sigma}) \qty(1 + \ln(1 \mp \frac{y - s}{\sigma}))^{\alpha+1} 
\nonumber\\
  & = \frac{f_Y(y)}{f_Y(s)} \qty(1 \mp \frac{y - s}{\sigma}) x^{-\qty(1+\alpha^{-1})} .
\end{align}

The pseudocode for a tail function using this mapping function is shown in Algorithm~\ref{alg:tail-exponential}.

\begin{algorithm}[H]
 \caption{Tail function with exponential mapping function}
 \label{alg:tail-exponential}
 \begin{algorithmic}[1]
 \Require $\alpha$ \Comment {Positive real satisfying Equation~\ref{eq:exponential-monotonicity-condition}}
 \Require $\sigma$ \Comment {Positive real satisfying Equation~\ref{eq:exponential-monotonicity-condition}}
 \Require \Call {PDF}{$x$} %\Comment {Cumulative Distribution Function}
 \Require \Call {RandReal}{\null} \Comment {Random real $\in [0,1)$}
 \Function{Tail}{$s$}\Comment{$s$ is the beginning of the tail distribution}
  \Repeat
   \State $u_1 \gets \Call{RandReal}{\null}$
   \State $t_1 \gets u_1^{-(\alpha^{-1})}$
   \State $t_2 \gets \exp(t_1 - 1)$
   \State $x \gets s \mp \sigma \qty(t_2 - 1)$
  \Until {$u_1 \times \Call{RandReal}{\null} < t_1 \times t_2 \times \frac{\Call {PDF}{x}}{\Call {PDF}{s}}$}
  \State \Return {$x$}
 \EndFunction
 \end{algorithmic}

\end{algorithm}

Again, the efficiency of rejection sampling can be calculated by integrating Equation~\ref{eq:exponential-accept-prob}. The 
result is
$ \frac{\alpha}{\sigma} \frac{F_Y(s)}{f_Y(s)}$
and
$ \frac{\alpha}{\sigma} \frac{\overline{F}_Y(s)}{f_Y(s)}$
for right and left tail distributions, respectively. The reader should not be deceived by the similarity of these expressions to
that of the previous
section into presuming they have the same rejection efficiency, since $\alpha$ and $\sigma$ must satisfy different constraint.
Generally the rejection efficiency of this method is inferior to those presented in 
Sections~\ref{sec:tail-logarithmic}, \ref{sec:tail-trigonometric}, and \ref{sec:tail-rational}
, however, it has a greater domain of applicability than those.

\subsection{Unbounded monotone densities}

\begin{figure}[H]
 \centering
 \input{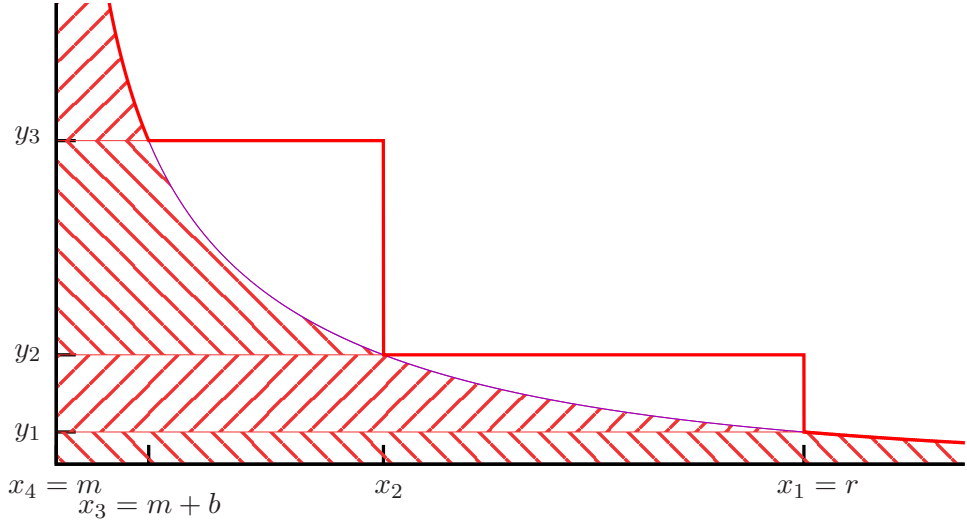}
 \caption{The partitioning scheme of generalized Ziggurat algorithm is applicable to unbounded densities as well.
 The $\chi_1^2$ distribution is shown here as an example. Hatched regions have the same area.}
 \label{fig:unbounded-pdf-new-partitions}
\end{figure}

As shown in Figure~\ref{fig:unbounded-pdf-new-partitions}, the partitioning scheme of generalized Ziggurat algorithm can handle
distributions with an unbounded density. This is because the total area under the PDF must be finite even if the density is not.
We will show here that a combination of nonlinear mapping and rejection sampling can be used here as well to generate random
numbers from such densities.

Let $m$ be the mode of the distribution where the density grows without bound: $\lim_{y \to m} f_Y(y) = \infty$.
We define the algebraic order of growth $q$ as the lowest $e$ that satisfies
\begin{equation}
 \label{eq:algebraic-order-of-growth}
 \lim_{e' \downarrow e} \qty(\lim_{y \to m} f_Y(y)\qty|y-m|^{e'}) \neq \infty .
\end{equation}
For example the algebraic order of growth for a gamma distribution with shape parameter $\alpha < 1$ is $1-\alpha$.
Note that the algebraic order of growth for a PDF is always a positive number less than 1, otherwise the CDF grows without bound.
If in addition to the limit in Equation~\ref{eq:algebraic-order-of-growth}, the following criteria holds as well
\begin{equation}
 \label{eq:algebraic-growth}
 \lim_{y \to m} f_Y(y)\qty|y-m|^q \neq \infty ,
\end{equation}
we say that $f_Y(y)$ has an algebraic growth. For example a PDF proportional to $\abs{\log(\abs{y-m})}$ near $m$, has an algebraic
order of growth 0, while it does not have an algebraic growth as it does not satisfy Equation~\ref{eq:algebraic-growth}.
Since all common and well-known unbounded distributions have an algebraic growth, we limit our treatment only to these kinds of 
PDFs.
This allows us to separate the PDF into an algebraic function responsible for the unbounded growth and a locally bounded
function:
\begin{equation}
 \label{eq:density-separation}
 f_Y(y) = \abs{y-m}^{-q} h_Y(y)
\end{equation}

We call the unbounded topmost region, the peak distribution, from now on.
Let $b$ be the width of the peak distribution's support such that the rightmost point on the support of the peak distribution 
will be $m+b$ if the density is a decreasing function.
Similarly if the density is an increasing function the leftmost point on the support of the peak distribution will be $m-b$.
For example $x_3 = m+b$ in the case of Figure~\ref{fig:unbounded-pdf-new-partitions}.
The density of the peak distribution is proportional to $f_Y(y)-f_Y(m \mp b)$.
Throughout this section we use the upper sign for increasing PDFs and the lower sign for decreasing PDFs (similar to
Figure~\ref{fig:unbounded-pdf-new-partitions}).

We use the following mapping function
\begin{equation}
 y = g(x) = m \mp b x^{\frac{1}{\beta}} ,
\end{equation}
where $\beta$ is a shape parameter between 0 and 1. Using Equation~\ref{eq:probability-change-of-var-2} but noting that the 
target density
is proportional to $f_Y(y)-f_Y(m \mp b)$, we find
\begin{equation}
 \label{eq:peak-density-propto}
 f_X(x) \propto \frac{f_Y(y)-f_Y(m \mp b)}{\abs{ \dv{g^{-1}(y)}{y} }}
 = \frac{f_Y(y)-f_Y(m \mp b)}{\frac{\beta}{b} \qty(\mp \frac{y-m}{b})^{\beta-1}} .
\end{equation}
Substituting Equation~\ref{eq:density-separation} into Equation~\ref{eq:peak-density-propto} results in
\begin{align}
 \label{eq:peak-accept-prob-propto}
 f_X(x) &\propto h_Y(y) \qty(\mp \frac{y-m}{b})^{1-\beta-q} - h_Y(m \mp b) \qty(\mp \frac{y-m}{b})^{1-\beta} \nonumber\\
        &= h_Y(y) x^{\frac{1-\beta-q}{\beta}} - h_Y(m \mp b) x^{\frac{1-\beta}{\beta}} .
\end{align}
In order for this density to be bounded, $\beta \leq 1-q$ must be satisfied.
Since $h_Y(y)$ is a locally bounded function, it must have a maxima in the support of peak distribution which will be
denoted by $h_{max}$. Also let $h_b$ denote $h_Y(m \mp b)$.
It can be shown that the maxima of the right hand side of the Equation~\ref{eq:peak-accept-prob-propto} is always less than
\begin{equation}
 \label{eq:peak-proportionality-const}
 A = \frac{h_b q}{1-\beta-q} \qty(\frac{1-\beta-q}{1-\beta})^{\frac{1-\beta}{q}} + h_{max} - h_b.
\end{equation}
The reciprocal of $A$ can serve as proportionality constant for
Equation~\ref{eq:peak-accept-prob-propto} to get the acceptance probability
\begin{align}
 \label{eq:peak-accept-prob}
 \Pr(x) & = \frac{1}{A} \qty(h_Y(y) x^{\frac{1-\beta-q}{\beta}} - h_Y(m \mp b) x^{\frac{1-\beta}{\beta}}) \nonumber\\
        & = \frac{b^q}{A} \qty\Big(f_Y(y) - f_Y(m \mp b)) \qty(\mp \frac{y-m}{b x}).
\end{align}
Assuming a constant $h_Y(y)$
\footnote{As usually a high number of blocks is employed in the Ziggurat algorithm, $b$ will be very small compared to
length scales of the distribution, and therefore for most practical purposes $h_Y(y)$ could be treated as a constant.}
, efficiency of rejection sampling can be estimated by integrating $\Pr(x)$ as
\begin{equation}
 \beta \frac{1-q-\beta}{1-q} \qty(\frac{1-\beta}{1-\beta-q})^{\frac{1-\beta}{q}} .
\end{equation}

We still need to choose $\beta$. Any value satisfying $0 < \beta \leq 1-q$ will do, so $\beta$ can be used to optimize the
efficiency. The optimal value of $\beta$ is a nonlinear function of $q$ and not easily computable, but
$\beta = \frac{1}{2} \qty(1 - q^2)$
approximates the optimal solution and gives an acceptable efficiency of more than $\frac{1}{2}$ for all values of $q$.
Substituting this into Equation~\ref{eq:peak-proportionality-const} results in
\begin{equation}
 A = 2 h_b q \frac{\qty(1-q)^{\frac{\qty(1-q)^2}{q}}}{\qty(1+q^2)^{\frac{1+q^2}{2 q}}} + h_{max} - h_b.
\end{equation}

Finally, this is the pseudocode for generating a number from an unbounded peak distribution:
\begin{algorithm}[H]
 \caption{function for generating random numbers from unbounded peak distributions}
 \label{alg:unbounded-peak}
 \begin{algorithmic}[1]
 \Require $m$ \Comment {mode}
 \Require $q$ \Comment {algebraic order of growth}
 \Require $h_{max}$ \Comment {see the text}
 \Require $h_b$ \Comment {see the text}
 \Require \Call {PDF}{$x$} %\Comment {Cumulative Distribution Function}
 \Require \Call {RandReal}{\null} \Comment {Random real $\in [0,1)$}
 \Function{UnboundedPeak}{$b$}
  \Statex\Comment{$b$ is the width of peak distribution's support}
  \State $E \gets \frac{2}{1-q^2}$
  \State $A \gets 2 h_b q \frac{\qty(1-q)^{\frac{\qty(1-q)^2}{q}}}{\qty(1+q^2)^{\frac{1+q^2}{2 q}}} + h_{max} - h_b$
  \State $C \gets \flatfrac{b^q}{A}$
  \Repeat
   \State $u_1 \gets \Call{RandReal}{\null}$
   \State $t \gets u_1^E$
   \State $x \gets m \mp b t$
  \Until {$u_1 \times \Call{RandReal}{\null} < C \times t \times \qty\Big(\Call {PDF}{x} - \Call {PDF}{m \mp b})$}
  \State \Return {$x$}
 \EndFunction
 \end{algorithmic}

\end{algorithm}

\subsection{Asymmetric distributions}

\begin{figure}[H]
 \centering
 \input{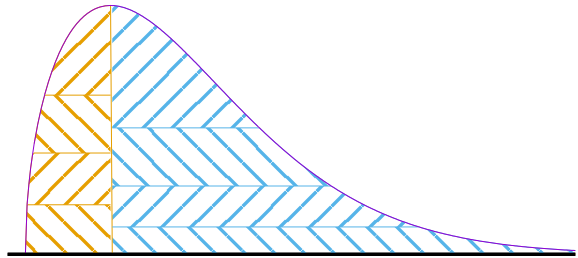}
 \caption{The partitioning scheme of generalized Ziggurat algorithm for asymmetric distributions.
 Shown here for Weibull distribution ($\alpha=1.5$) with 4 regions. Hatched regions of the same color have the same area.}
 \label{fig:asymmetric-pdf-partitions}
\end{figure}

Asymmetric unimodal distributions are separated into two monotonic distributions on either side of the mode.
Their partitioning will be independent of each other. Figure~\ref{fig:asymmetric-pdf-partitions} shows the partitioning scheme for
the Weibull distribution as an example. The ratio of each distribution's area to the total area is computed.
Each time a random number is generated, one of the distributions is randomly selected with a probability equal to its ratio of
area and a random number will be returned from that distribution.

\subsection{Ziggurat setup}

\begin{figure}[H]
 \centering
 \input{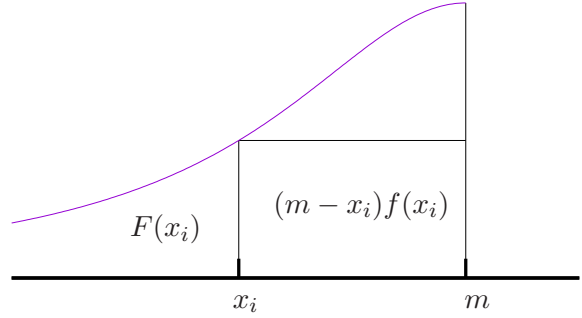}
 \caption{The area of a strip of height $f(x_i)$}
 \label{fig:pdf-area}
\end{figure}

In this section we show how coordinates satisfying requirements of new partitioning scheme, could be found.
As shown in Figure~\ref{fig:pdf-area}, for an increasing density function, the area of a strip with coordinate $x_i$ is
$F(x_i) + (m-x_i) f(x_i)$. Similarly, for a decreasing density function it would be $\overline{F}(x_i) + (x_i-m)f(x_i)$.
Therefore, we define the following area function:
\begin{equation}
 A(x) =
 \begin{dcases}
  F(x) + (m-x) f(x) & \text{ for increasing PDFs}
  \\
  \overline{F}(x) + (x-m)f(x) & \text{ for decreasing PDFs}
 \end{dcases}
\end{equation}
Note that since we're dealing with a monotonic density function, $A(x)$ would also be monotonic.

If the total area under the PDF is $A_{tot}$ and we would like to partition the distribution into $N$ regions,
the first (bottommost) strip should have an area equal to $\frac{A_{tot}}{N}$, and the sum of area of the first two regions
should be $2\frac{A_{tot}}{N}$, and so on. In other words:
\begin{equation}
 A(x_i) = i \frac{A_{tot}}{N} \quad\text{ for } i \text{ from 1 to } N
\end{equation}
Generally, the inverse of $A(x)$ is not easily computable, but since $A(x)$ is monotonic, this nonlinear equation could be solved
using the bisection method. We define the following residue functions $R_i(x)=A(x) - i \frac{A_{tot}}{N}$ whose roots are the
$x_i$ coordinates:
\begin{equation}
 R_i(x_i) = 0 \quad\text{ for } i \text{ from 1 to } N .
\end{equation}

The bisection method initially needs two points whose residues have opposite signs. Since $A(x)$ is monotonic, it will be equal to
0 and $A_{tot}$ at the endpoints of the density support. So the residue functions at the endpoints of the density support will
have opposite signs for all $i \neq N$. The case of $i=N$ is trivial: $x_N = m$.

For PDFs with bounded support, starting from the endpoints of support is fine and straightforward, but for PDFs with unbounded 
support, starting from numerical infinity will be slow and wasteful as most of the coordinates will be close to the mode anyway.
We use a method similar to multiplicative binary search to find an interval where the residue function has opposing signs.
We first consider some neighbourhood of the mode, if the residue function doesn't have opposing signs in there, we shift and 
expand the
interval away from the mode. Algorithm~\ref{alg:find-interval} shows the pseudocode for this.

\begin{algorithm}[H]
 \caption{function for finding an initial interval for the bisection method}
 \label{alg:find-interval}
 \begin{algorithmic}[1]
 \Require $m$ \Comment {mode}
 \Require \Call {Residue}{$x$}
 \Function{FindInterval}{\null}
  \State $l \gets 1$
  \State $a \gets m$
  \State $b \gets m \mp l$
  \While {$\Call{Residue}{a} \times \Call{Residue}{b} > 0$}
   \State $a \gets b$
   \State $l \gets 2 \times l$
   \State $b \gets m \mp l$
  \EndWhile
  \State \Return {$a$, $b$}
 \EndFunction
 \end{algorithmic}
\end{algorithm}
Here (as before) the upper sign is used for increasing PDFs, and the lower sign is used for decreasing PDFs. The typical length
scale of the PDF could also be used as the initial value for $l$, but that would affect the generality of the algorithm.

\section{Canonical floating-point random number generator}\label{sec:canonic-float-rng}

As described in the introduction the result of division of a random integer by its range is a fixed-point number which unlike
a floating-point number does not enjoy increased precision near 0. When such random numbers are used in the tail algorithm they
cause premature termination of the tail and large gaps between produced random numbers near the termination point.
We generate floating-point random numbers with fully random fraction bits for the tail algorithm.
To generate such a random floating-point number, we use a uniform random number as the fraction and a geometric random number
as the exponent, as suggested by \citet{gaussian-survey}.

In a uniform random integer each bit is either zero or one with independent equal probability $\frac{1}{2}$.
Therefore, the position of the first non-zero bit $l_1$ follows a geometric distribution.
An $n$-bit integer could be zero with probability $2^{-n}$.
In this case we need to generate a new random integer and add $n$ to the $l_1$ of the new number.
This procedure should be repeated until a non-zero number is found.
Let's suppose we have a uniform
random bit generator (URBG) producing $b$-bit random integers and we would like to produce floating-point numbers 
with $f$ fraction (explicit mantissa) bits and also assume $b > f$.
\footnote{If $b \leq f$, then the URBG should be called $n = \lceil \frac{f}{b} \rceil$ times.}
$f$ bits of the random integer is multiplied by $2^{-f}$ and the implicit 1 is added to get a random number $m$ in $[1,2)$ with 
fully
random fraction, and the remaining $b-f$ bits are used to generate a geometric random number $g$.
The desired random floating-point number is $m \times 2^{-g}$
The pseudocode for this is shown in Algorithm~\ref{alg:canonic-float-rng}

\begin{algorithm}[H]
 \caption{function for generating random floating-point numbers with fully random fraction}
 \label{alg:canonic-float-rng}
 \begin{algorithmic}[1]
 \Require \Call {Ran}{\null} \Comment {URBG producing $b$-bit integers}
 \Require $b$ \Comment {number of bits in the random integer}
 \Require $f$ \Comment {number of bits in the fraction of a floating-point}
 \Function{CanonicalRandReal}{\null}
  \State $U \gets \Call{Ran}{\null}$
  \State $m \gets 1 + 2^{-f} \times$ ($f$ MSBs of $U$)
  \State $r \gets (b-f)$ LSBs of $U$
  \State $g \gets 1$
  \If {$r = 0$}
   \State $g \gets g + b -f$
   \State $r \gets \Call{Ran}{\null}$
   \While {$r = 0$}
    \State $g \gets g + b$
    \State $r \gets \Call{Ran}{\null}$
   \EndWhile
  \EndIf
  \Loop
   \If {rightmost bit of $r$ = 1} \Return {$m \times 2^{-g}$}
   \EndIf
   \State Shift $r$ one bit to the right
   \State $g \gets g + 1$
  \EndLoop
 \EndFunction
 \end{algorithmic}
\end{algorithm}

Of course it does not matter whether we find the first 0 bit or the first 1 bit, or whether we count from the left or the right.
So the required geometric exponent could be efficiently calculated using assembly instructions such as count leading zeros (clz), 
count trailing zeros (ctz), count leading ones (clo), find first set (ffs), and bit scan reverse (bsr). Most CPU architectures 
provide at least one of these instructions and almost all modern Intel and AMD CPUs have the clz instruction (as \code{LZCNT}).
Most compilers also provide similar intrinsics. With the aid of these instructions, most of the time the geometric exponent could 
be calculated in just one or two CPU cycles.
However we have abandoned their use in favour of a portable and \proglang{C++} standard conforming implementation.

Another way to optimize Algorithm~\ref{alg:canonic-float-rng} is to precompute and store the geometric exponents.
Note that the probability of executing the if block in lines 6-13 is $2^{f-b}$
which is less than 0.2\% when generating single
precision floating-point numbers from 32-bit integers and less than 0.025\% when generating double precision floating-point
numbers from 64-bit integers.
When it is not executed, $r$ has a value in $[1,2^{b-f}-1]$, and therefore all the corresponding
$2^{-g}$ values could be precomputed which takes only 2 KiB and 16 KiB of space for single and double precision cases, 
respectively.
Therefore, most of the time, we could merely multiply $m$ by these cached values, as shown in 
Algorithm~\ref{alg:optimized-canonic-float-rng}.

\begin{algorithm}[H]
 \caption{Optimized version of Algorithm~\ref{alg:canonic-float-rng}}
 \label{alg:optimized-canonic-float-rng}
 \begin{algorithmic}[1]
 \Require \Call {Ran}{\null} \Comment {URBG producing $b$-bit integers}
 \Require $b$ \Comment {number of bits in the random integer}
 \Require $f$ \Comment {number of bits in the fraction of a floating-point}
 \Require $multiplier[1..2^{b-f}]$ \Comment {cached values explained above}
 \Function{CanonicRandReal}{\null}
  \State $U \gets \Call{Ran}{\null}$
  \State $m \gets 1 + 2^{-f} \times$ ($f$ MSBs of $U$)
  \State $r \gets (b-f)$ LSBs of $U$
  \State $g \gets 1$
  \If {$r \neq 0$} \Return{$m \times multiplier[r]$}
  \Else
   \State $g \gets g + b -f$
   \State $r \gets \Call{Ran}{\null}$
   \While {$r = 0$}
    \State $g \gets g + b$
    \State $r \gets \Call{Ran}{\null}$
   \EndWhile
  \EndIf
  \Loop
   \If {rightmost bit of $r$ = 1} \Return {$m \times 2^{-g}$}
   \EndIf
   \State Shift $r$ one bit to the right
   \State $g \gets g + 1$
  \EndLoop
 \EndFunction
 \end{algorithmic}
\end{algorithm}

Compared to merely dividing a random integer by its range, this algorithm only costs an extra multiply-add, two bit mask, one
non-zero check, and an array lookup most of the time. These costs are negligible compared to a random number generation using a 
modern URBG such as Mersenne Twister.

\section{Implementation details}\label{sec:implementation}

In Sections~\ref{sec:original-algorithm} and \ref{sec:generalized-algorithm}, we omitted some common optimizations for the sake 
of clarity.
In this section we describe those that can be used to improve performance as well as those that should \emph{not} be used
because they affect the quality of generated random numbers. We also describe which tail algorithm is used for each of the
implemented distributions.

Usually a random floating-point number is generated by multiplying a random integer in $[0,2^b-1]$ by $2^{-b}$, where $b$ is the
number of bits in the random integer (often 32 or 64). In Ziggurat algorithm these random numbers must be multiplied by $x_i$
to map them to $[0,x_i$). One multiplication can be saved by precomputing a table of $2^{-b} x_i$ values. Moreover, instead of
comparing the random floating-point number - which is in $[0,x_i)$ - with $x_{i+1}$, the random integer can be compared against
a precomputed integer table of $2^b \frac{x_{i+1}}{x_i}$. This replaces the floating-point comparison with integer comparison.
However, as the speed difference is marginal in modern platforms, it is not utilized in our implementation.
Applying both of these modifications makes the table of $x_i$ coordinates redundant and therefore it can be deleted.
When dealing with symmetric distributions, instead of generating a random sign,
signed random numbers can be directly produced by reinterpreting the unsigned integer as signed and replacing $b$ by $b-1$ in the
above table.

Many Ziggurat implementations select the random region using the least significant bits of the same random integer that was used 
to produce the random floating-point number. As mentioned in the introduction and explained by \citet{doornik}, this makes
random numbers correlated. To avoid this we mask out (effectively set them to zero) those bits used as an index to
select region before using the integer to produce a random floating-point number.

\section{Usage instructions}\label{sec:usage}

\pkg{Zest} is a single header template library, therefore its usage is very simple. First of all it should be included and
imported into the global namespace:

\begin{verbatim}
#include <random>
#include "zest.hpp"
using namespace zest;
\end{verbatim}

Then a random engine and a Ziggurat object should be constructed. This constructs a Mersenne Twister random engine and a Ziggurat
for a standard normal distribution:

\begin{verbatim}
std::mt19937_64 urbg;
Ziggurat<StandardNormal, std::mt19937_64> ziggurat_for_std_normal;
\end{verbatim}

Then each time a random number is needed, \code{ziggurat\_for\_normal(urbg)} should simply be called.

The Ziggurat object can be constructed in a similar manner for other standard distributions and chi-squared distribution.

\begin{verbatim}
Ziggurat<StandardExponential, std::mt19937_64> ziggurat_for_std_exponential;

Ziggurat<StandardCauchy, std::mt19937_64> ziggurat_for_std_cauchy;

Ziggurat<ChiSquared<3>, std::mt19937_64> ziggurat_for_chi_sq_w_3_dof;
\end{verbatim}

All the other distributions need extra parameters and a distribution object must be constructed first and then passed to 
the Ziggurat constructor:

\begin{verbatim}
Weibull weibull_dist {2.5, 3};
Ziggurat<Weibull, std::mt19937_64> ziggurat_for_weibull {weibull_dist};
\end{verbatim}

The order, meaning, and the default values of constructor's arguments are identical to those of their standard library
counterparts. This is a list of their prototypes here for a quick reference:

\begin{verbatim}
Normal (double mean = 0.0, double stddev = 1.0);
Cauchy (double mode = 0.0, double scale = 1.0);
Exponential (double rate = 1.0);
Gamma (double shape = 1.0, double scale = 1.0);
Weibull (double shape = 1.0, double scale = 1.0);
LogNormal (double normal_mean = 0.0, double normal_stddev = 1.0);
StudentT (double dof);
FisherF (double dof1 = 1.0, double dof2 = 1.0);
\end{verbatim}

For more advanced usage and guidelines on how to define your custom PDF, please read the README file.

\section{Results and discussion}\label{sec:results}

In order to ensure that the distribution of generated numbers accurately represents the theoretical distribution
all probability distributions are evaluated using the Kolmogorov-Smirnov test
\citep{kstest1,art}.
This test measures the maximum vertical distance $D_N$ between a specified continuous distribution function and empirical
distribution function obtained from a sample of $N$ random number. The Kolmogorov distribution is the probability distribution of
$D_N$ assuming the sample is drawn from the hypothesized continuous distribution (the so-called null hypothesis). If the $p$~value
which is the probability of observing deviations at least as large as $D_N$ is negligible, the null hypothesis is rejected and
the alternative hypothesis that the sample came from a different distribution is accepted. If the $p$~value is small but
non-negligible (e.g. between 0.01 and 0.1), the null hypothesis is considered suspicious, and further tests needs to be carried
out. Otherwise the null hypothesis is accepted as no justification for its rejection is found.

It is important to note that the $p$~value is the probability of observing the measured deviation conditional on the null 
hypothesis
being true, and does not represent the probability of the null hypothesis being true conditional on observing such deviation,
because in general $\Pr(A|B) \neq \Pr(B|A)$. Moreover, in any one test any deviation can possibly happen with its corresponding
probability. Equivalently it could be said that the resulting $p$~value should have a uniform distribution.
So repeating the test $M$ times and testing the $M$ $p$~values for uniformity makes the overall test much stronger.
As noted by \citet[p. 52]{art} this method ``tends to detect both local and global nonrandom behavior''.

We used $M=2^{10}$ samples, each with $N=2^{20}$ numbers. The $p$~value of each sample is computed using the asymptotic formulas
as $N$ is large enough to permit an accurate calculation. The final uniformity test's $p$~value is calculated using the exact 
method and the code developed by \citet{kolmogorov-calc}. All distributions passed the test.

To compare the performance of \pkg{Zest} with that of \pkg{Boost} \citep{boost} and \pkg{Standard Template Library (STL)} 
\citep{stl},
the time needed to generate $2^{26}$ double precision floating-point numbers is measured.
This procedure is repeated $2^4$ times with different seeds and the results are averaged.
The generalized Ziggurat algorithm is tested with 256, 1024, and 4092 regions. The 64-bit version of the MT19937 random number
generator \citep{mt19937,mt19937_64} is employed.
The average time needed to produce one random variate in nanoseconds will be presented in the following.
Reported uncertainties are the standard error of the mean.

\subsection{Normal distribution}

\pkg{STL} and \pkg{Boost} implement the Box-Muller and the original Ziggurat (with 256 regions) algorithms, respectively.
\pkg{Zest} employs the IPDF tail algorithm to generate random numbers from the normal distribution's tail.

\begin{table}[H]
\resizebox{.99\textwidth}{!}{
\begin{tabular}{c c c c c}
 \toprule
 \pkg{Zest} ($N=256$) & \pkg{Zest} ($N=1024$) & \pkg{Zest} ($N=4092$) & \pkg{STL} & \pkg{Boost}
 \\ \midrule
 $\SI{12.564 \pm 0.015}{\ns}$ &
$\SI{12.1231 \pm 0.0092}{\ns}$ &
$\SI{12.103 \pm 0.010}{\ns}$ &
$\SI{43.516 \pm 0.025}{\ns}$ &
$\SI{12.667 \pm 0.020}{\ns}$

 \\ \bottomrule
\end{tabular}
}
\caption{Normal variate generation times (in nanoseconds)}
\label{tab:normal-time}
\end{table}

As can be seen from Table~\ref{tab:normal-time}, both Ziggurat implementations have similar performance and both are more than 3.6
times faster than the \pkg{STL}'s Box-Muller algorithm.

\subsection{Cauchy distribution}

\pkg{STL} and \pkg{Boost} both transform a uniform variate using the Cauchy ICDF.
\pkg{Zest} employs the ICCDF tail algorithm to generate random numbers from the Cauchy distribution's tail.

\begin{table}[H]
\resizebox{.99\textwidth}{!}{
\begin{tabular}{c c c c c}
 \toprule
 \pkg{Zest} ($N=256$) & \pkg{Zest} ($N=1024$) & \pkg{Zest} ($N=4092$) & \pkg{STL} & \pkg{Boost}
 \\ \midrule
 $\SI{13.348 \pm 0.018}{\ns}$ &
$\SI{12.599 \pm 0.021}{\ns}$ &
$\SI{12.465 \pm 0.016}{\ns}$ &
$\SI{54.911 \pm 0.043}{\ns}$ &
$\SI{51.220 \pm 0.036}{\ns}$

 \\ \bottomrule
\end{tabular}
}
\caption{Cauchy variate generation times (in nanoseconds)}
\label{tab:cauchy-time}
\end{table}

The observed times in Table~\ref{tab:cauchy-time} indicates that \pkg{Zest} is about 4 times faster than both \pkg{STL} and 
\pkg{Boost}.

\subsection{Exponential distribution}

\pkg{STL} transforms a uniform variate using the exponential ICDF.
\pkg{Boost} implements the original Ziggurat algorithm. By taking advantage of the self similarity of the exponential 
distribution,
\pkg{Boost} generates random numbers from the tail distribution by shifting its Ziggurat.
Table~\ref{tab:exponential-time-initial} shows that \pkg{Zest} is faster than \pkg{STL}, but is slower than \pkg{Boost}.
\pkg{Boost}'s performance could be explained by its tail algorithm and its use of special pretests (squeeze improvements) 
specifically designed for the exponential distribution.

\begin{table}[H]
\resizebox{.99\textwidth}{!}{
\begin{tabular}{c c c c c}
 \toprule
 \pkg{Zest} ($N=256$) & \pkg{Zest} ($N=1024$) & \pkg{Zest} ($N=4092$) & \pkg{STL} & \pkg{Boost}
 \\ \midrule
 $\SI{17.955 \pm 0.018}{\ns}$ &
$\SI{16.859 \pm 0.028}{\ns}$ &
$\SI{16.640 \pm 0.027}{\ns}$ &
$\SI{42.684 \pm 0.036}{\ns}$ &
$\SI{11.279 \pm 0.054}{\ns}$

 \\ \bottomrule
\end{tabular}
}
\caption{Exponential variate generation times when \pkg{Zest} employs the ICCDF tail algorithm (in nanoseconds)}
\label{tab:exponential-time-initial}
\end{table}

\subsection{Gamma distribution}

\pkg{STL} implements an algorithm by \citet{marsaglia-gamma}, while \pkg{Boost} implements an algorithm by 
\citet{ahrens-gamma}.
\pkg{Zest} uses the logarithmic tail function for the gamma distribution. For a gamma distribution with shape parameter $\alpha$ 
and
scale parameter $\theta$, the optimal value of $\sigma$ in Equation~\ref{eq:logarithmic-monotonicity-condition} is $\theta$ when 
$\alpha \leq 1$ and $\theta \frac{s}{s - (\alpha-1)\theta}$ when $\alpha > 1$.

\begin{table}[H]
\resizebox{.99\textwidth}{!}{
\begin{tabular}{c c c c c c}
 \toprule
 $\alpha$ & \pkg{Zest} ($N=256$) & \pkg{Zest} ($N=1024$) & \pkg{Zest} ($N=4092$) & \pkg{STL} & \pkg{Boost}
 \\ \midrule
 0.1 & $\SI{93.038 \pm 0.067}{\ns}$ &
$\SI{38.543 \pm 0.042}{\ns}$ &
$\SI{23.728 \pm 0.017}{\ns}$ &
$\SI{179.14 \pm 0.11}{\ns}$ &
$\SI{90.660 \pm 0.075}{\ns}$
 \\
 0.2 & $\SI{35.677 \pm 0.019}{\ns}$ &
$\SI{23.332 \pm 0.012}{\ns}$ &
$\SI{19.698 \pm 0.018}{\ns}$ &
$\SI{178.189 \pm 0.099}{\ns}$ &
$\SI{99.606 \pm 0.039}{\ns}$
 \\
 0.5 & $\SI{25.762 \pm 0.027}{\ns}$ &
$\SI{20.276 \pm 0.015}{\ns}$ &
$\SI{18.726 \pm 0.010}{\ns}$ &
$\SI{109.887 \pm 0.081}{\ns}$ &
$\SI{112.771 \pm 0.033}{\ns}$
 \\
 1 & $\SI{19.568 \pm 0.014}{\ns}$ &
$\SI{18.3076 \pm 0.0093}{\ns}$ &
$\SI{18.085 \pm 0.012}{\ns}$ &
$\SI{89.179 \pm 0.080}{\ns}$ &
$\SI{11.174 \pm 0.013}{\ns}$
 \\
 2.5 & $\SI{33.156 \pm 0.030}{\ns}$ &
$\SI{31.533 \pm 0.022}{\ns}$ &
$\SI{31.305 \pm 0.027}{\ns}$ &
$\SI{85.736 \pm 0.038}{\ns}$ &
$\SI{230.30 \pm 0.36}{\ns}$
 \\
 10 & $\SI{34.911 \pm 0.034}{\ns}$ &
$\SI{33.349 \pm 0.033}{\ns}$ &
$\SI{33.116 \pm 0.031}{\ns}$ &
$\SI{84.985 \pm 0.032}{\ns}$ &
$\SI{234.96 \pm 0.13}{\ns}$
 \\
 100 & $\SI{35.978 \pm 0.022}{\ns}$ &
$\SI{33.998 \pm 0.021}{\ns}$ &
$\SI{33.721 \pm 0.030}{\ns}$ &
$\SI{84.121 \pm 0.036}{\ns}$ &
$\SI{243.17 \pm 0.15}{\ns}$

 \\ \bottomrule
\end{tabular}
}
\caption{Gamma variate generation times (in nanoseconds)}
\label{tab:gamma-time}
\end{table}

As can be seen from Table~\ref{tab:gamma-time}, for $\alpha>1$ \pkg{Zest} is about 2.5 and 7 times faster than \pkg{STL} and 
\pkg{Boost}, respectively. \pkg{Boost} recognizes the special case of $\alpha=1$ as the exponential distribution and treats it 
accordingly, 
which 
explains its performance for this case. \pkg{Zest} is also at least twice faster than both \pkg{STL} and \pkg{Boost} for $0.1 < 
\alpha < 1$ when at least 1024 regions is used, but the 
rejection efficiency drops as $\alpha$ is lowered, especially for lower number of regions. This decrease in rejection efficiency 
is not due to the tail or the unbounded peak algorithm, but rather due to low rejection efficiency of finite regions.

\subsection{Chi-squared distribution}

Chi-squared distribution is a special case of gamma distribution, and therefore both \pkg{STL} and \pkg{Boost} use their gamma 
generators to
generate chi-squared variates. \pkg{Zest} also uses tail algorithm similar to the one used for gamma variates, but the special 
cases of 
1 and 2 degrees of freedom are handled differently. To generate random numbers from the unbounded peak of $\chi_1^2$ 
distribution, a value of $\beta=\frac{1}{2}$ is used. This reduces the rejection efficiency but replaces a floating-point power 
evaluation with a multiplication in the evaluation of $g(x)$ as $\frac{1}{\beta}=2$. The $\chi_2^2$ distribution is treated as an 
exponential distribution.

\begin{table}[H]
\resizebox{.99\textwidth}{!}{
\begin{tabular}{c c c c c c}
 \toprule
 $k$ & \pkg{Zest} ($N=256$) & \pkg{Zest} ($N=1024$) & \pkg{Zest} ($N=4092$) & \pkg{STL} & \pkg{Boost}
 \\ \midrule
 1 & $\SI{19.9880 \pm 0.0096}{\ns}$ &
$\SI{17.607 \pm 0.012}{\ns}$ &
$\SI{16.993 \pm 0.014}{\ns}$ &
$\SI{110.89 \pm 0.16}{\ns}$ &
$\SI{112.034 \pm 0.055}{\ns}$
 \\ \bottomrule
\end{tabular}
}
\caption{Chi-Squared variate generation times (in nanoseconds)}
\label{tab:chi-squared-time}
\end{table}

Table~\ref{tab:chi-squared-time} shows that $\chi_1^2$ variate generation is slightly faster than generating gamma variates with 
$\alpha=0.5$
in \pkg{Zest}, which is due to the use of $\beta=\frac{1}{2}$.

\subsection{Weibull distribution}

Both \pkg{STL} and \pkg{Boost} use the Weibull's ICDF to transform uniform variates into Weibull variates. \pkg{Zest} uses the 
Weibull's ICCDF to
generate random numbers from the tail distribution.

\begin{table}[H]
\resizebox{.99\textwidth}{!}{
\begin{tabular}{c c c c c c}
 \toprule
 $\alpha$ & \pkg{Zest} ($N=256$) & \pkg{Zest} ($N=1024$) & \pkg{Zest} ($N=4092$) & \pkg{STL} & \pkg{Boost}
 \\ \midrule
 0.1 & $\SI{160.60 \pm 0.31}{\ns}$ &
$\SI{56.414 \pm 0.032}{\ns}$ &
$\SI{28.522 \pm 0.021}{\ns}$ &
$\SI{127.674 \pm 0.060}{\ns}$ &
$\SI{125.986 \pm 0.058}{\ns}$
 \\
 0.2 & $\SI{50.913 \pm 0.039}{\ns}$ &
$\SI{27.737 \pm 0.034}{\ns}$ &
$\SI{20.857 \pm 0.017}{\ns}$ &
$\SI{117.547 \pm 0.049}{\ns}$ &
$\SI{116.272 \pm 0.065}{\ns}$
 \\
 0.5 & $\SI{31.076 \pm 0.026}{\ns}$ &
$\SI{21.818 \pm 0.012}{\ns}$ &
$\SI{19.168 \pm 0.016}{\ns}$ &
$\SI{43.730 \pm 0.030}{\ns}$ &
$\SI{41.970 \pm 0.034}{\ns}$
 \\
 1 & $\SI{19.715 \pm 0.016}{\ns}$ &
$\SI{18.3074 \pm 0.0096}{\ns}$ &
$\SI{18.110 \pm 0.015}{\ns}$ &
$\SI{43.791 \pm 0.039}{\ns}$ &
$\SI{42.047 \pm 0.017}{\ns}$
 \\
 2.5 & $\SI{36.078 \pm 0.041}{\ns}$ &
$\SI{33.940 \pm 0.025}{\ns}$ &
$\SI{33.428 \pm 0.034}{\ns}$ &
$\SI{116.268 \pm 0.083}{\ns}$ &
$\SI{115.189 \pm 0.042}{\ns}$
 \\
 10 & $\SI{36.066 \pm 0.032}{\ns}$ &
$\SI{33.523 \pm 0.035}{\ns}$ &
$\SI{32.966 \pm 0.033}{\ns}$ &
$\SI{116.29 \pm 0.20}{\ns}$ &
$\SI{115.56 \pm 0.50}{\ns}$
 \\
 100 & $\SI{39.989 \pm 0.033}{\ns}$ &
$\SI{33.911 \pm 0.021}{\ns}$ &
$\SI{32.665 \pm 0.040}{\ns}$ &
$\SI{115.860 \pm 0.056}{\ns}$ &
$\SI{114.524 \pm 0.029}{\ns}$

 \\ \bottomrule
\end{tabular}
}
\caption{Weibull variate generation times (in nanoseconds)}
\label{tab:weibull-time}
\end{table}

Table~\ref{tab:weibull-time} presents the generation time for Weibull variates. The results shows that \pkg{Zest} is more than 
3 times faster
than both \pkg{STL} and \pkg{Boost} for $\alpha>1$. It's also at least twice faster than both \pkg{STL} and \pkg{Boost} for $0.1 
< \alpha < 1$ when 
$N=1024$. The cause of decreased rejection efficiency is the same as that of the gamma distribution.
The special case of $\alpha=1$ is the exponential distribution and the case of $\alpha=0.5$ replaces a floating-point power 
evaluation with a square root, which explains \pkg{STL} and \pkg{Boost} performance at these two values.

\subsection{Log-normal distribution}

Both \pkg{STL} and \pkg{Boost} generate log-normal variates by exponentiating the result of their normal generators. 
\pkg{Zest} uses the IIPDF tail algorithm with $d(x)=\frac{1}{x}$ and $\alpha = \frac{\sigma^2}{s-\mu}$ to generate random numbers 
from the tail distribution
Table~\ref{tab:log-normal-time-initial} shows the generation times. The results shows that \pkg{Zest} is faster than both 
\pkg{STL} and \pkg{Boost}. It should be noted that log-normal distributions with a large $\sigma$ have extremely narrow and high 
peaks along with extremely slow decaying tails which reduces the rejection efficiency, as can be seen in the case of $\sigma=5$.

\begin{table}[H]
\resizebox{.99\textwidth}{!}{
\begin{tabular}{c c c c c c c}
 \toprule
 $\mu$ & $\sigma$ & \pkg{Zest} ($N=256$) & \pkg{Zest} ($N=1024$) & \pkg{Zest} ($N=4092$) & \pkg{STL} & \pkg{Boost}
 \\ \midrule
 0 & 0.2 & $\SI{34.535 \pm 0.037}{\ns}$ &
$\SI{32.655 \pm 0.035}{\ns}$ &
$\SI{32.283 \pm 0.027}{\ns}$ &
$\SI{71.696 \pm 0.068}{\ns}$ &
$\SI{38.939 \pm 0.090}{\ns}$
 \\
 0 & 1 & $\SI{31.723 \pm 0.032}{\ns}$ &
$\SI{29.289 \pm 0.024}{\ns}$ &
$\SI{28.715 \pm 0.030}{\ns}$ &
$\SI{70.646 \pm 0.032}{\ns}$ &
$\SI{37.448 \pm 0.017}{\ns}$
 \\
 0 & 5 & $\SI{58.313 \pm 0.050}{\ns}$ &
$\SI{32.715 \pm 0.033}{\ns}$ &
$\SI{28.043 \pm 0.040}{\ns}$ &
$\SI{71.750 \pm 0.048}{\ns}$ &
$\SI{38.850 \pm 0.026}{\ns}$
 \\
 -10 & 1 & $\SI{31.612 \pm 0.026}{\ns}$ &
$\SI{29.316 \pm 0.029}{\ns}$ &
$\SI{28.728 \pm 0.031}{\ns}$ &
$\SI{70.487 \pm 0.059}{\ns}$ &
$\SI{37.275 \pm 0.052}{\ns}$
 \\
 10 & 1 & $\SI{31.653 \pm 0.031}{\ns}$ &
$\SI{29.307 \pm 0.031}{\ns}$ &
$\SI{28.689 \pm 0.030}{\ns}$ &
$\SI{70.457 \pm 0.035}{\ns}$ &
$\SI{37.285 \pm 0.025}{\ns}$

 \\ \bottomrule
\end{tabular}
}
\caption{Log-normal variate generation times when \pkg{Zest} employs the asymmetric algorithm}
\label{tab:log-normal-time-initial}
\end{table}

\subsection{Student's t distribution}

A student's t variate with $\nu$ degree of freedom can be generated from a standard normal variate $\mathcal{N}$, and a 
chi-squared variate $\chi_{\nu}^2$ as $\frac{\mathcal{N}}{\nu \sqrt{\chi_{\nu}^2}}$. Both \pkg{STL} and \pkg{Boost} use this 
method. \pkg{Zest} 
generates student's t variates directly. The IPDF tail algorithm is employed with $\alpha=\frac{\nu+1}{\nu}$ as described in 
Section~\ref{sec:tail-ipdf}. The results presented in Table~\ref{tab:student-time} show that \pkg{Zest} is 10 times faster than 
the \pkg{STL} and 20 
times faster than \pkg{Boost} for $\nu > 1$. For $\nu < 1$ the rejection efficiency of the tail region and finite regions close 
to 
it starts to drop. When using 1024 regions a relative speedup of at least 5 for $\nu>0.2$ could still be maintained.
Using more regions improves the \pkg{Zest}'s performance at lower values of $\nu$.

\begin{table}[H]
\resizebox{.99\textwidth}{!}{
\begin{tabular}{c c c c c c}
 \toprule
 $\nu$ & \pkg{Zest} ($N=256$) & \pkg{Zest} ($N=1024$) & \pkg{Zest} ($N=4092$) & \pkg{STL} & \pkg{Boost}
 \\ \midrule
 0.1 & $\SI{112.244 \pm 0.064}{\ns}$ &
$\SI{38.730 \pm 0.047}{\ns}$ &
$\SI{19.213 \pm 0.017}{\ns}$ &
$\SI{226.56 \pm 0.20}{\ns}$ &
$\SI{99.373 \pm 0.047}{\ns}$
 \\
 0.2 & $\SI{28.288 \pm 0.020}{\ns}$ &
$\SI{16.805 \pm 0.014}{\ns}$ &
$\SI{13.4415 \pm 0.0079}{\ns}$ &
$\SI{225.74 \pm 0.17}{\ns}$ &
$\SI{105.141 \pm 0.059}{\ns}$
 \\
 0.5 & $\SI{18.059 \pm 0.015}{\ns}$ &
$\SI{13.764 \pm 0.015}{\ns}$ &
$\SI{12.577 \pm 0.012}{\ns}$ &
$\SI{224.47 \pm 0.16}{\ns}$ &
$\SI{112.314 \pm 0.056}{\ns}$
 \\
 1 & $\SI{13.409 \pm 0.021}{\ns}$ &
$\SI{12.274 \pm 0.013}{\ns}$ &
$\SI{12.1062 \pm 0.0088}{\ns}$ &
$\SI{156.11 \pm 0.15}{\ns}$ &
$\SI{126.763 \pm 0.089}{\ns}$
 \\
 2.5 & $\SI{14.230 \pm 0.031}{\ns}$ &
$\SI{12.510 \pm 0.011}{\ns}$ &
$\SI{12.255 \pm 0.011}{\ns}$ &
$\SI{132.896 \pm 0.077}{\ns}$ &
$\SI{255.72 \pm 0.15}{\ns}$
 \\
 10 & $\SI{13.530 \pm 0.011}{\ns}$ &
$\SI{12.328 \pm 0.013}{\ns}$ &
$\SI{12.210 \pm 0.016}{\ns}$ &
$\SI{129.715 \pm 0.074}{\ns}$ &
$\SI{239.20 \pm 0.14}{\ns}$
 \\
 100 & $\SI{13.336 \pm 0.014}{\ns}$ &
$\SI{12.261 \pm 0.011}{\ns}$ &
$\SI{12.193 \pm 0.012}{\ns}$ &
$\SI{128.401 \pm 0.098}{\ns}$ &
$\SI{256.27 \pm 0.16}{\ns}$

 \\ \bottomrule
\end{tabular}
}
\caption{Student's t-distribution variate generation times (in nanoseconds)}
\label{tab:student-time}
\end{table}

\subsection{Fisher's f distribution}

A Fisher's f variate with $d_1$ and $d_2$ degrees of freedom could be generated from two chi-squared variates $\chi_{d_1}^2$ and 
$\chi_{d_2}^2$ as $\flatfrac{\frac{\chi_{d_1}^2}{d_1}}{\frac{\chi_{d_2}^2}{d_2}}$. Both \pkg{STL} and \pkg{Boost} use this method.
\pkg{Zest} uses the rational tail algorithm with $\alpha=\frac{d_2}{2}$ and $\sigma = s + \frac{d_2}{d_1}\frac{d_1+d_2}{d_2+2}$ 
for 
$d_1 < 2$ and $\sigma = s + \frac{s d_2 (d_1 + d_2)}{s d_1 (d_2 + 2) - d_2 (d_1 -2)}$ for $d_1 > 2$ as the optimal values 
satisfying Equation~\ref{eq:rational-monotonicity-condition}.

\begin{table}[H]
\resizebox{.99\textwidth}{!}{
\begin{tabular}{c c c c c c c}
 \toprule
 $d_1$ & $d_2$ & \pkg{Zest} ($N=256$) & \pkg{Zest} ($N=1024$) & \pkg{Zest} ($N=4092$) & \pkg{STL} & \pkg{Boost}
 \\ \midrule
 0.2 & 0.2 & $\SI{426.28 \pm 0.23}{\ns}$ &
$\SI{110.948 \pm 0.065}{\ns}$ &
$\SI{41.487 \pm 0.027}{\ns}$ &
$\SI{361.82 \pm 0.23}{\ns}$ &
$\SI{182.294 \pm 0.070}{\ns}$ \\
 0.5 & 0.5 & $\SI{52.329 \pm 0.016}{\ns}$ &
$\SI{31.801 \pm 0.017}{\ns}$ &
$\SI{23.751 \pm 0.021}{\ns}$ &
$\SI{358.61 \pm 0.19}{\ns}$ &
$\SI{197.58 \pm 0.18}{\ns}$ \\
 1 & 1 & $\SI{34.809 \pm 0.024}{\ns}$ &
$\SI{23.113 \pm 0.015}{\ns}$ &
$\SI{19.491 \pm 0.016}{\ns}$ &
$\SI{225.136 \pm 0.096}{\ns}$ &
$\SI{226.85 \pm 0.15}{\ns}$ \\
 2 & 2 & $\SI{20.636 \pm 0.012}{\ns}$ &
$\SI{18.679 \pm 0.021}{\ns}$ &
$\SI{18.1886 \pm 0.0100}{\ns}$ &
$\SI{180.881 \pm 0.053}{\ns}$ &
$\SI{23.826 \pm 0.029}{\ns}$ \\
 10 & 10 & $\SI{34.086 \pm 0.037}{\ns}$ &
$\SI{31.412 \pm 0.016}{\ns}$ &
$\SI{30.804 \pm 0.019}{\ns}$ &
$\SI{172.913 \pm 0.068}{\ns}$ &
$\SI{446.54 \pm 0.26}{\ns}$ \\
 100 & 100 & $\SI{35.964 \pm 0.020}{\ns}$ &
$\SI{33.621 \pm 0.035}{\ns}$ &
$\SI{33.101 \pm 0.025}{\ns}$ &
$\SI{170.13 \pm 0.13}{\ns}$ &
$\SI{480.12 \pm 0.55}{\ns}$ \\
 0.2 & 100 & $\SI{116.500 \pm 0.061}{\ns}$ &
$\SI{44.525 \pm 0.028}{\ns}$ &
$\SI{25.020 \pm 0.014}{\ns}$ &
$\SI{271.88 \pm 0.17}{\ns}$ &
$\SI{333.24 \pm 0.32}{\ns}$ \\
 1 & 100 & $\SI{27.958 \pm 0.015}{\ns}$ &
$\SI{20.644 \pm 0.034}{\ns}$ &
$\SI{18.537 \pm 0.012}{\ns}$ &
$\SI{200.70 \pm 0.16}{\ns}$ &
$\SI{356.99 \pm 0.24}{\ns}$ \\
 2 & 100 & $\SI{21.036 \pm 0.026}{\ns}$ &
$\SI{18.752 \pm 0.014}{\ns}$ &
$\SI{18.224 \pm 0.013}{\ns}$ &
$\SI{178.29 \pm 0.11}{\ns}$ &
$\SI{253.474 \pm 0.098}{\ns}$ \\
 10 & 100 & $\SI{34.901 \pm 0.020}{\ns}$ &
$\SI{32.667 \pm 0.025}{\ns}$ &
$\SI{32.146 \pm 0.024}{\ns}$ &
$\SI{173.656 \pm 0.097}{\ns}$ &
$\SI{464.11 \pm 0.69}{\ns}$ \\
 100 & 0.2 & $\SI{164.030 \pm 0.080}{\ns}$ &
$\SI{63.707 \pm 0.057}{\ns}$ &
$\SI{36.848 \pm 0.032}{\ns}$ &
$\SI{271.09 \pm 0.12}{\ns}$ &
$\SI{333.58 \pm 0.17}{\ns}$ \\
 100 & 1 & $\SI{38.485 \pm 0.034}{\ns}$ &
$\SI{30.971 \pm 0.043}{\ns}$ &
$\SI{28.765 \pm 0.024}{\ns}$ &
$\SI{202.3 \pm 1.7}{\ns}$ &
$\SI{357.31 \pm 0.24}{\ns}$ \\
 100 & 2 & $\SI{32.969 \pm 0.029}{\ns}$ &
$\SI{29.802 \pm 0.046}{\ns}$ &
$\SI{28.948 \pm 0.029}{\ns}$ &
$\SI{178.35 \pm 0.11}{\ns}$ &
$\SI{253.28 \pm 0.15}{\ns}$ \\
 100 & 10 & $\SI{34.457 \pm 0.022}{\ns}$ &
$\SI{31.684 \pm 0.029}{\ns}$ &
$\SI{31.062 \pm 0.039}{\ns}$ &
$\SI{173.490 \pm 0.093}{\ns}$ &
$\SI{466.1 \pm 1.9}{\ns}$
 \\ \bottomrule
\end{tabular}
}
\caption{Fisher's F-distribution variate generation times (in nanoseconds)}
\label{tab:fisher-time}
\end{table}

Table~\ref{tab:fisher-time} shows that \pkg{Zest} is 5 to 10 faster than both \pkg{STL} and \pkg{Boost} when $d_1,d_2 \geq 1$. 
Rejection efficiency starts to suffer
however when either of $d_1$ or $d_2$ is less than 2 and when either of $d_1$ or $d_2$ is less than or equal to 0.2, 
number of regions should be 1024 or more to preserve overall efficiency.

When either of $d_1$ or $d_2$ equals 2, one of the required chi-squared variates is an exponential variates. This explains
\pkg{Boost}'s good performance for $d_1=2,d_2=2$ case where both of them can be generated using fast exponential generators.

\newpage
\section{Summary} \label{sec:summary}

The generalized Ziggurat algorithm can efficiently generate random variates from unimodal distributions with unbounded support 
and/or unbounded density. Several algorithms for generating variates from tail distributions has been presented. Their 
preconditions and efficiencies has been discussed. In general the tail algorithm whose covering distribution produces a 
tighter fit around the original distribution should be used. So logarithmic tail algorithm (when applicable) is preferable to 
trigonometric and rational tail algorithm which are in turn preferable to the exponential tail algorithm in terms of rejection 
efficiency.

The \pkg{Zest} library is presented which can be used to generate normal, log-normal, exponential, chi-squared, gamma, Weibull, 
Cauchy, student's t, and Fisher's f distribution efficiently.
Our results shows that \pkg{Zest} is faster than both \pkg{STL} and \pkg{Boost}, except for the 
case of unbounded densities with very small shape parameters, and the exponential distribution (only slower than \pkg{Boost}). 
Ziggurat algorithm has a high setup time, so it's not suitable 
for applications that require variates with frequently changing shape parameters.
\pkg{Zest} is publicly available at \url{https://github.com/DiscreteLogarithm/Zest}.

An efficient algorithm and implementation is also presented which can generate true floating-point random numbers. It is capable 
of producing all representable floating-point values in [0,1). This implementation is only about 25\% slower than the divide by 
range method in the single-precision floats case,
while in the case of double-precision floats it is even about 25\% faster. This is due to the fact multiplying an integer with 64 
significant bits by a floating-point number requires a quad precision intermediate, while this algorithm uses 
an integer with 52 bits to generate the fraction.

\section*{Computational details}

All tests are carried out on a linux computer (kernel Version~4.6.11) with Intel core i7-2640M (2.8 GHz) CPU and GCC compiler
(Version~7.3.1). The \pkg{STL} \citep{stl} implementation used is that of the GCC (of the same version). Version~1.67.0 of 
\pkg{Boost} \citep{boost} is used.

\bibliography{refs}

\end{document}